\documentclass[reprint,superscriptaddress,amsmath,amssymb,showpacs,floatfix]{revtex4-2}
\usepackage[latin1]{inputenc}
\usepackage{amsfonts}
\usepackage{amssymb}
\usepackage{calrsfs}
\usepackage{amsmath}
\usepackage[dvipsnames]{xcolor}
\usepackage{graphicx}
\usepackage[colorlinks=true,allcolors=blue]{hyperref}
\hypersetup{breaklinks=true}

\begin{document}

	\title{Thermodynamic anomalies in overdamped systems with time-dependent temperature}    
	\author{Shakul Awasthi}
	\affiliation{School of Physics, Korea Institute for Advanced Study, Seoul 02455, Korea}
	\author{Hyunggyu Park}
	\affiliation{Quantum Universe Center, Korea Institute for Advanced Study, Seoul 02455, Korea}
	\author{Jae Sung Lee}
	\email{jslee@kias.re.kr}
	\affiliation{School of Physics, Korea Institute for Advanced Study, Seoul 02455, Korea}
	\affiliation{Quantum Universe Center, Korea Institute for Advanced Study, Seoul 02455, Korea}
	
	\date{\today}
	
\begin{abstract}
    One of the key objectives in investigating small stochastic systems is the development of micrometer-sized engines and the understanding of their thermodynamics. However, the primary mathematical tool used for this purpose, the overdamped approximation, has a critical limitation: it fails to fully capture the thermodynamics when the temperature varies over time, as the velocity is not considered in the approximation.
    Specifically, we show that heat dissipation and entropy production calculated under the overdamped approximation deviate from their true values. These discrepancies are termed thermodynamic anomalies. 
    To overcome this limitation, we analytically derive expressions for these anomalies in the presence of a general time-varying temperature.
    One notable feature of the result is that high viscosity and small mass, though both leading to the same overdamped dynamic equations, result in different thermodynamic anomaly relations. 
    Our results have significant implications, particularly for accurately calculating the efficiency of heat engines operating in overdamped environments with time-varying temperatures, without requiring velocity measurements. Additionally, our findings offer a simple method for estimating the kinetic energy of an overdamped system.
\end{abstract}
\maketitle
	

\section*{Introduction}  

Recent studies on stochastic systems have driven significant advancements in thermodynamics at the microscopic scale. These advancements have enabled the miniaturization of heat engines to a microscopic level~\cite{Whalen2003,Blickle2012,Steeneken2011,Brantut2013,Argun2017}, particularly using a single colloidal particle in overdamped environments~\cite{Blickle2012, Martinez2016,Krishnamurthy2016,Krishnamurthy2023}. The dynamics of such small systems are often analyzed by considering only their position trajectories while neglecting velocity variables. This simplification is justified because experimental systems are typically overdamped, meaning that velocity relaxes to equilibrium much faster than position dynamics in environments with high viscosity or negligible inertia effects. Due to its simplicity in mathematical handling, this overdamped approximation is widely employed to describe the dynamics observed in mesoscopic experiments involving typical fluids. In addition, this approximation is, in some sense, inevitable, as accurately measuring velocity in overdamped systems is challenging due to its rapid relaxation.

However, the overdamped approximation does not always guarantee accurate estimation of thermodynamic quantities, such as heat and entropy production (EP), even in environments with high viscosity or small mass. One example is a system with a position-dependent temperature~\cite{Widder1989,Hondou2000,Celani2012,Bo2013,Kawaguchi2013,Polettini2013,Sancho2015,Marino2016}. In such systems, EP calculated under the overdamped approximation differs from that obtained using the full underdamped formalism, which explicitly accounts for the velocity variable~\cite{Celani2012}. The difference between the two EPs is attributed to the symmetry breaking of time and velocity. As shown in this example, the finite discrepancy between thermodynamic quantities calculated under the overdamped approximation and the underdamped formulation is referred to as a \emph{thermodynamic anomaly}~\cite{Celani2012}. 

Therefore, thermodynamic anomalies play a crucial role in calculating thermodynamic quantities when velocity measurement is challenging. Despite their importance, there has been no systematic study on thermodynamic anomalies induced by time-dependent temperature, aside from several reports on specific systems~\cite{Schmiedl2007_2,Arold2019,Awasthi2022}. Investigating these anomalies in systems with time-varying temperature is essential due to their broad applicability in both theoretical and experimental contexts.
This is particularly relevant in the field of microscopic heat engines, where temperature varies periodically over time. Indeed, numerous microscopic heat engine models have been proposed over the past two decades~\cite{Parrondo1996,Derenyi1999,Van2005,Schmiedl2007_2,Rana2014,Park2016,Rossnagel2016,Martinez2017,Lee2019,Holubec2020_2,Chen2022,Majumdar2022}. 
If these anomalies can be properly evaluated and accounted for, thermodynamic quantities related to such engines might be accurately calculated using the simpler overdamped formalism, without requiring velocity measurements.

	

Here, we explicitly calculate the thermodynamic anomalies in heat and EP for systems with general time-varying temperature. Our results demonstrate that the two conditions, high viscosity and small mass, result in different thermodynamic anomaly relations, though both leading to the same overdamped dynamics. 
Moreover, we also find general anomaly relations in between the two conditions by introducing the two-parameter Brinkman's hierarchy method.
This indicates that accurately estimating thermodynamic quantities in overdamped systems with time-varying temperatures requires understanding the underlying mechanism that leads to the overdamped regime. Through numerical examples, we demonstrate that heat, efficiency of a heat engine, and kinetic energy can be accurately estimated in overdamped environments without the need for intricate velocity measurements, even when the temperature varies relatively quickly.

	%
	%
	\section*{Results}\label{Sec:Results}
	\subsection*{Setup}
	Consider a one-dimensional Brownian particle of mass $m$ immersed in a reservoir with a time-varying temperature $T(t)$. The position and velocity of the particle at time $t$ are denoted by $x_t$ and $v_t$, respectively. The motion of the particle is described by a stochastic differential equation known as the underdamped Langevin equation, given by
	\begin{equation} \label{lang}
		\dot{x}_t = v_t,~~	m \dot{v_t} = f(x_t,\lambda_t) - \gamma v_t + \eta_t~,
	\end{equation} 
	where $f(x_t,\lambda_t)$ denotes an external force applied on the particle, $\lambda_t$ represents a given time-dependent protocol, and $\gamma$ is the viscous coefficient. $\eta_t$ denotes the thermal Gaussian-white noise characterized by a zero mean and the autocorrelation $\langle \eta_t \eta_{t'} \rangle = 2 \gamma T(t) \delta(t-t')$ with the Boltzmann constant set to $k_{\rm B} = 1$. 
	Both conservative and nonconservative forces can be included in $f(x_t,\lambda_t)$. 
	The probability distribution $P_{\rm ud}(x,v,t)$ for the stochastic variables $x_t$ and $v_t$ at time $t$ is governed by the following Fokker-Planck (FP) equation~\cite{Risken1996}:
	\begin{equation}\label{FP-ud}
		\partial_t P_\text{ud}(x,v,t) = \mathcal{L}_{\rm ud} P_\text{ud}(x,v,t)~,
	\end{equation}
	where the underdamped FP operator~$\mathcal{L}_{\rm ud}$ is defined as
	\begin{equation}\label{FP-op}
		\mathcal{L}_\text{ud} :=  -\partial_x v - \frac{1}{m}\partial_v \left[f(x,\lambda) -\gamma v - \frac{\gamma T(t)}{m}  \partial_v \right]  ~.
	\end{equation}
	
	We now turn our attention to the thermodynamics of the system. According to the first law of thermodynamics at the trajectory level~\cite{Sekimoto1998}, the mean value of heat rate $\dot Q$ in the underdamped system is given by
	\begin{equation} \label{eq:Q_ud_1}
		\langle \dot Q \rangle_\text{ud} = -\langle \gamma v^2 \rangle_\text{ud} + \langle v \circ \eta\rangle_\text{ud}~,
	\end{equation}
	where the symbol $\circ$ denotes the Stratonovich product and $\langle \cdots \rangle_\text{ud}$ represents the ensemble average taken over the probability distribution $P_\text{ud}(x,v,t)$. The expression of the heat rate in Eq.~\eqref{eq:Q_ud_1} is equivalent to 
	\begin{equation} \label{eq:Q_ud_2}
		\langle \dot Q \rangle_\text{ud} = \int_{-\infty}^{\infty}\,dx \int_{-\infty}^{\infty}\,dv~m v~ J_\text{ud}^\text{irr} (x,v,t)~,  
	\end{equation} 
	where $J_\text{ud}^\text{irr}(x,v,t)$ is defined as
	\begin{equation}
		J_\text{ud}^\text{irr}(x,v,t) \equiv \left( - \frac{\gamma v}{m} - \frac{\gamma T(t)}{m^2} \partial_v \right) P_\text{ud}(x,v,t)~. 
	\end{equation}
	Then, the rate of total EP can be expressed using $J_\text{ud}^\text{irr}(x,v,t)$ as follows~\cite{Lee2023,Dechant2018,Spinney2012,Kwon2024}: 
	\begin{equation} \label{eq:EP_rate_ud}
		\langle \dot S_\text{tot} \rangle_\text{ud} = \int_{-\infty}^{\infty}\,dx \int_{-\infty}^{\infty}\,dv~ \frac{(m J_\text{ud}^\text{irr})^2}{\gamma T(t) P_\text{ud}}~.
	\end{equation}

	\subsection*{Time scales and overdamped equations}
	There are four characteristic time scales in this setup: (i) the velocity relaxation time $\tau_{\rm r} \equiv m/\gamma$, (ii) the time interval between two consecutive observations of the system $\tau_{\rm obs}$, (iii) the time scale of temperature variation $\tau_{\rm tmp}$, and (iv) the typical time scale of $x$-variable (overdamped) dynamics $\tau_{\rm od}$.
	When the condition $ \tau_{\rm r}/\tau_{\rm obs} \ll 1$ is satisfied, the velocity is always relaxed to equilibrium for any observation time.
	If the temperature is time-independent, the overdamped description using only the $x$ variable is valid under the single condition $ \tau_{\rm r}/\tau_{\rm obs} \ll 1$. However, if the temperature is time-dependent, the overdamped approximation also depends on $\tau_{\rm tmp}$: for example, when $\tau_{\rm tmp} \approx \tau_{\rm r}$ (indicating very rapid variation of temperature), a proper overdamped description cannot be obtained, as explained in the Supplementary Section V~\cite{supp_NC}. Thus, in this study, we consider the following hierarchy of time scales:  $\tau_{\rm r} \ll \tau_{\rm obs} \leq \tau_{\rm od} \approx \tau_{\rm tmp}$.
	
	On these time scales, the systematic overdamped approximation can be carried out using Brinkman's hierarchy method~\cite{Brinkman1956,Risken1996}. In this method, there are two scaling parameters: $m$ and $\gamma$. To satisfy the condition $\tau_{\rm r} \ll \tau_{\rm obs}$, either a small $m$ or a large $\gamma$ is required. 
	Since these two regimes can lead to distinct approximated expressions for various thermodynamic quantities, it is important to carefully examine the scaling behaviors of the parameters. To systematically carry out this investigation, we introduce two different dimensionless scaling parameters for Brinkman's hierarchy method, enabling the exploration of more general scaling behaviors, including small $m$ and large $\gamma$ regimes.
	As elaborated in the Methods section, the two parameters are defined as $\tau \equiv \tau_{\rm od}/\tau_{\rm r} \gg 1$ and $\nu \equiv V_{\rm ud}/V_{\rm od}$, where $V_{\rm ud}$ and $V_{\rm od}$ are the characteristic velocities in underdamped and overdamped dynamics, respectively. These parameters are related by $\nu = \tau^z$ $(0\leq z \leq 1/2)$. Here, $z=0$ and $z=1/2$ correspond to the large $\gamma$ and small $m$ limits, respectively, while $0< z < 1/2$ represents an intermediate regime between the two limiting cases and may correspond to a specific experimental setup. 
	
	Through this method, we first derive the overdamped approximation of the underdamped FP equation~\eqref{FP-ud}. The detailed derivation is presented in the Methods section. 
	Regardless of the $z$ value, the resulting equation is the same as the usual overdamped FP equation, as shown below:
	\begin{equation}\label{FP-od}
		\partial_t P_\text{od}(x,t) = -\partial_x J_{\rm od}(x,t), 
	\end{equation}
	where $P_\text{od}(x,t) = \int dv P_\text{ud}(x,v,t)$ and $J_\text{od}(x,t) \equiv \gamma^{-1} \left( f(x,t)  - T(t) \partial_x \right) P_\text{od}(x,t)$. Therefore, the corresponding overdamped Langevin equation is 
	\begin{equation}\label{over-lang}
		\gamma \dot{x_t} = f(x_t,\lambda_t) + \eta_t~,
	\end{equation} 
	which is identical to the expression obtained by simply neglecting the inertia term in Eq.~\eqref{lang}.

	%
	%
	
	\subsection*{Thermodynamic anomalies}
	Conventionally, the mean heat rate of the overdamped equation~\eqref{over-lang} is known as~\cite{Sekimoto1998} 
	\begin{equation} \label{eq:heat_rate_over_conv}
		\langle \dot Q \rangle_\text{od} = -\langle f(x,\lambda) \circ \dot{x} \rangle_{\rm od} = -\int_{-\infty}^{\infty}\,dx f(x,\lambda) J_\text{od}(x,t),
	\end{equation}
	where $\langle \cdots \rangle_\text{od}$ denotes the ensemble average taken over $P_\text{od}(x,t)$. 
	Additionally, the rate of total EP in the overdamped approximation is conventionally expressed as~\cite{Seifert2005,Seifert2012}
	\begin{equation} \label{eq:EP_rate_over_conv}
		\langle \dot S_\text{tot} \rangle_\text{od} = \int_{-\infty}^{\infty}\,dx~ \frac{\gamma J_\text{od}^2}{ T(t) P_\text{od}}~.
	\end{equation}
	These two expressions, Eqs.~\eqref{eq:heat_rate_over_conv} and \eqref{eq:EP_rate_over_conv}, accurately quantify heat and EP in overdamped dynamics when the temperature is time-independent. 
	
	However, if the temperature has time dependence, the overdamped approximations of Eqs.~\eqref{eq:Q_ud_2} and \eqref{eq:EP_rate_ud} do not conincide with Eqs.~\eqref{eq:heat_rate_over_conv} and \eqref{eq:EP_rate_over_conv}, respectively. We refer to this discrepancy as the thermodynamic anomaly. Our main result is the explicit expressions for these anomalies. 
	First, the heat anomaly, defined as $\mathcal{A}_Q \equiv \langle \dot Q\rangle_{\rm ud} - \langle \dot Q\rangle_{\rm od}$ up to the same order of $\langle \dot Q\rangle_{\rm od}$, is
	\begin{equation}\label{dQ-ud}
		\mathcal{A}_Q = 
		\begin{cases} 
			\frac{\dot{T}}{2} - \frac{m \ddot{T}}{4 \gamma} & \text{for }~ z=0\;, \\ 
			\frac{\dot{T}}{2} & \text{for }~ 0 < z \leq 1/2\;.
		\end{cases}
	\end{equation}
	Therefore, the anomaly depends on $z$, unlike the dynamic equation. 
	The orders of the terms $\dot{T}/2$ and $m \ddot{T}/4\gamma$ are $O(\tau^{0})$ and $O(\tau^{-1})$, respectively, and are independent of $z$. In contrast, the order of $\langle \dot Q\rangle_{\rm od}$ depends on $z$ and is given by $O(\tau^{-1+2z})$.
	Therefore, for $z=0$ (high-viscosity regime), among the three terms contributing to $\langle \dot Q\rangle_{\rm ud}$, $\dot{T}/2$ is the leading-order term, while the other two terms, $\langle \dot Q\rangle_{\rm od}$ and $m \ddot{T}/4\gamma$, are of the same higher order $O(\tau^{-1})$. For $0<z < 1/2$, $\langle \dot Q\rangle_{\rm od}$ is of higher order than $\dot{T}/2$. Finally, for $z=1/2$ (small-mass regime), $\dot{T}/2$ and $\langle \dot Q\rangle_{\rm od}$ are of the same order, $O(\tau^{0})$. 
	Note that the order of each term mentioned above is estimated using the dimensionless formalism presented in the Methods section.
	An interesting feature of $\mathcal{A}_Q$ is that the time-accumulated $\mathcal{A}_Q$ depends only on the initial and final information of the temperature, not on the stochastic path, even though heat is not a state variable. Furthermore, $\mathcal{A}_Q$ is independent of the external force applied to the system.

	The heat anomaly arises from neglecting the velocity degree of freedom in the overdamped approximation. This approximation implicitly assumes that the velocity is always relaxed to its equilibrium state. 
	Consequently, for constant temperature, the kinetic energy $E_{\rm K}$ remains unchanged, resulting in no additional heat exchange with the environment. However, when the temperature varies with time, $E_{\rm K}$ also changes, leading to additional heat exchange with the environment. This is the origin of the heat anomaly. As demonstrated in Supplementary Section IV~\cite{supp_NC}, an explicit relation between $\mathcal{A}_Q$ and $E_{\rm K}$ is given by 
	\begin{equation} \label{eq:kineticE}
		\langle \dot E_{\rm K} \rangle = \mathcal{A}_Q
		{~~\rm or~~}
		\langle E_{\rm K}\rangle_{\rm ud} = 
		\begin{cases} 
			\frac{T}{2}  - \frac{m \dot{T}}{4 \gamma} & \text{for } z=0\;, \\ 
			\frac{T}{2} & \text{for } 0 < z \leq 1/2\;.
		\end{cases}
	\end{equation}
	up to the order of $\langle \dot Q\rangle_{\rm od}$.

	Similarly to the case of heat, the total EP exhibits a discrepancy between Eq.~\eqref{eq:EP_rate_ud} and Eq.~\eqref{eq:EP_rate_over_conv}. The EP anomaly, defined as $\mathcal{A}_S \equiv \langle \dot S_\text{tot} \rangle_\text{ud} - \langle \dot S_\text{tot} \rangle_\text{od}$ up to the same order of $\langle \dot S_\text{tot} \rangle_\text{od}$, is
	\begin{equation}\label{ent-prod-z}
		\mathcal{A}_S = 
		\begin{cases} 
			\frac{m}{4 \gamma} \left(\frac{\dot{T}}{T}\right)^2 & \text{if } z=0 \;,\\ 
			0 & \text{if } 0 < z \leq 1/2\;.
		\end{cases}
	\end{equation}
	The details of the calculations are provided in Supplementary Section III~\cite{supp_NC}.
	Thus, $\mathcal{A}_S$ also depends on $z$. The order of $\frac{m}{4 \gamma} (\dot{T}/T)^2$ is $O(\tau^{-1})$ and is independent of $z$, whereas the order of $\langle \dot S_\text{tot} \rangle_\text{od}$ is $O(\tau^{-1+2z})$. 
	Therefore, for $z=0$, the orders of
	$\langle \dot S_\text{tot} \rangle_\text{od}$ and $\frac{m}{4 \gamma} (\dot{T}/T)^2$ are the same, $O(\tau^{-1})$. For $0<z\leq 1/2$, notably, no EP anomaly appears in this range of $z$, even though $\mathcal{A}_Q$ remains finite.
	Unlike $\mathcal{A}_Q$, the time-accumulated $\mathcal{A}_S$ is path-dependent for $z=0$ due to the squared term of the time derivative of the temperature. Moreover, Eq.~\eqref{ent-prod-z} indicates that $\langle \dot S_\text{tot} \rangle_\text{ud} \geq \langle \dot S_\text{tot} \rangle_\text{od}$. 
	Similar to the case of position-dependent temperature~\cite{Celani2012}, the finite $\mathcal{A}_S$ for $z = 0$ originates from the breaking of time- and velocity-reversal symmetry at the microscopic level. However, for $0 < z \leq 1/2$, the extent of this symmetry breaking is negligible compared to $\langle \dot{S}_{\rm tot} \rangle_{\rm od}$, leading to the absence of the anomaly.

	%
	%

	\begin{figure}[t] 
		\includegraphics[width=0.95\columnwidth]{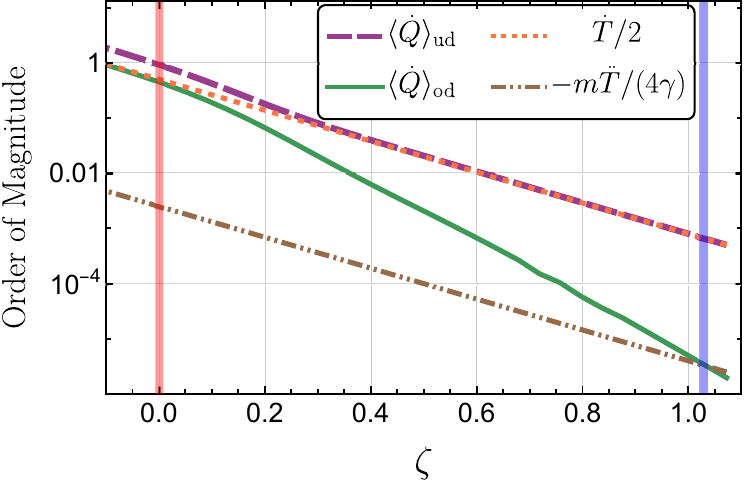}
		\caption{Plots of $\langle \dot Q\rangle_{\rm ud}$, $\langle \dot Q\rangle_{\rm od}$, $\dot T/ 2$, and $m\ddot T/(4\gamma)$ as functions of $\zeta$. 
			The $z$ values at the two crossing points - between $\langle \dot Q\rangle_{\rm od}$ and $\dot T/2$, and between $\langle \dot Q\rangle_{\rm od}$ and $m\ddot T/(4\gamma)$ - are highlighted by the orange and purple vertical lines, respectively. The orange and purple lines are located at $\zeta \approx 0$ and $\zeta \approx 1.03$, respectively.
			For each $\zeta$, the values of $\langle \dot Q\rangle_{\rm ud}$, $\langle \dot Q\rangle_{\rm od}$, $\dot T/2$, and $m\ddot T/(4\gamma)$ are taken as their maximum absolute values within one period in the periodic steady state.} 
		\label{Fig_1}
	\end{figure}

	\subsection*{Estimation and control of $z$}
	Equations~\eqref{dQ-ud} and \eqref{ent-prod-z} show that thermodynamic anomalies depend on the exponent $z$. Therefore, identifying or controlling $z$ in the given system is crucial for precisely estimating the anomalies in overdamped dynamics. 
	Here, we propose an experimental or simulation method to estimate or control the exponent $z$. Adjusting $z$ can be achieved by simultaneously varying the amplitudes of both the external force and the temperature. Consider the following adjustments to their magnitudes:
	\begin{equation} \label{eq:f_T_scaling}
		f \rightarrow \tau^{-\zeta}f~~~{\rm and}~~~T \rightarrow \tau^{-\zeta}T,\;
	\end{equation}
	where $\zeta$ is a control parameter for the magnitude adjustment. Then, it is straightforward to see that the dimensionless coefficient equation~\eqref{eqA:cbn-eqn_dimless} remains invariant under this magnitude control, except for the change from $\nu=\tau^z$ to $\nu^\prime \equiv \tau^{z-\zeta/2}$. This clearly indicates that the exponent $z$ can be controlled by adjusting $\zeta$, or equivalently, by varying the magnitudes of $f$ and $T$. 
	
	In practice, the exponent can be estimated or set to an appropriate value by comparing the magnitudes of $\dot T/2$, $\langle \dot Q\rangle_{\rm od}$, and $m\ddot T/(4\gamma)$. In the setup with the magnitude adjustment~\eqref{eq:f_T_scaling}, the orders of $\dot T/2$ and $m\ddot T/(4\gamma)$ are $O(\tau^{0})$ and $O(\tau^{-1})$, respectively, whereas the order of $\langle \dot Q\rangle_{\rm od}$ depends on $z-\zeta/2$ as $\langle \dot Q\rangle_{\rm od} \sim O(\tau^{-1+2(z-\zeta/2)})$. 
	Now, we can estimate $\dot T/2$, $\langle \dot Q\rangle_{\rm od}$, and $m\ddot T/(4\gamma)$ by varying $\zeta$, as shown in Fig.~\ref{Fig_1}. 
	If the magnitudes of $\langle \dot Q\rangle_{\rm od}$ and $m\ddot T/(4\gamma)$ become comparable, it implies that $-1 +2(z-\zeta/2) \approx -1$, which yields $z \approx \zeta/2$. On the other hand, if the magnitudes of $\langle \dot Q\rangle_{\rm od}$ and $\dot T/2$ become comparable, it signifies $-1 +2(z-\zeta/2) \approx 0$, which yields $z \approx \zeta/2 + 1/2$.

	To verify this method numerically, we consider a Langevin system with an external force $f=k_0 x$ and a time-dependent temperature $T(t) = 2 + \sin(t)$. Here, we set $\tau_{\rm tmp} = 2\pi = \tau_{\rm od}$, which is the period of the temperature variation. Thus, $\tau = 2\pi \gamma/m$. The parameters are set as $k_0 = 1$, $\gamma = 1$, and $m=0.01$ (small-$m$ condition). Figure~\ref{Fig_1} shows the plots of $\langle \dot Q\rangle_{\rm od}$, $\dot T/2$, and $m \ddot T/(4\gamma)$ as functions of $\zeta$. Comparing the magnitudes of $\langle \dot Q\rangle_{\rm od}$ with $m\ddot T/(4\gamma)$ or $\dot T/2$ consistently leads to $z\approx 1/2$, which corresponds to the small-$m$ setup.
	
	We note that the exponent can also be extracted from the scaling behavior of $\langle \dot Q \rangle_{\rm od} \sim \tau^{-1+2z}$ by varying $\tau$. However, this approach may not be suitable, as adjusting $\tau$ requires varying $m$ or $\gamma$, which are typically fixed parameters in most experimental setups rather than controllable variables. In contrast, adjusting the magnitudes of $f$ and $T$ is more straightforward.

    \begin{figure}[t]
	\includegraphics[width=0.95\linewidth]{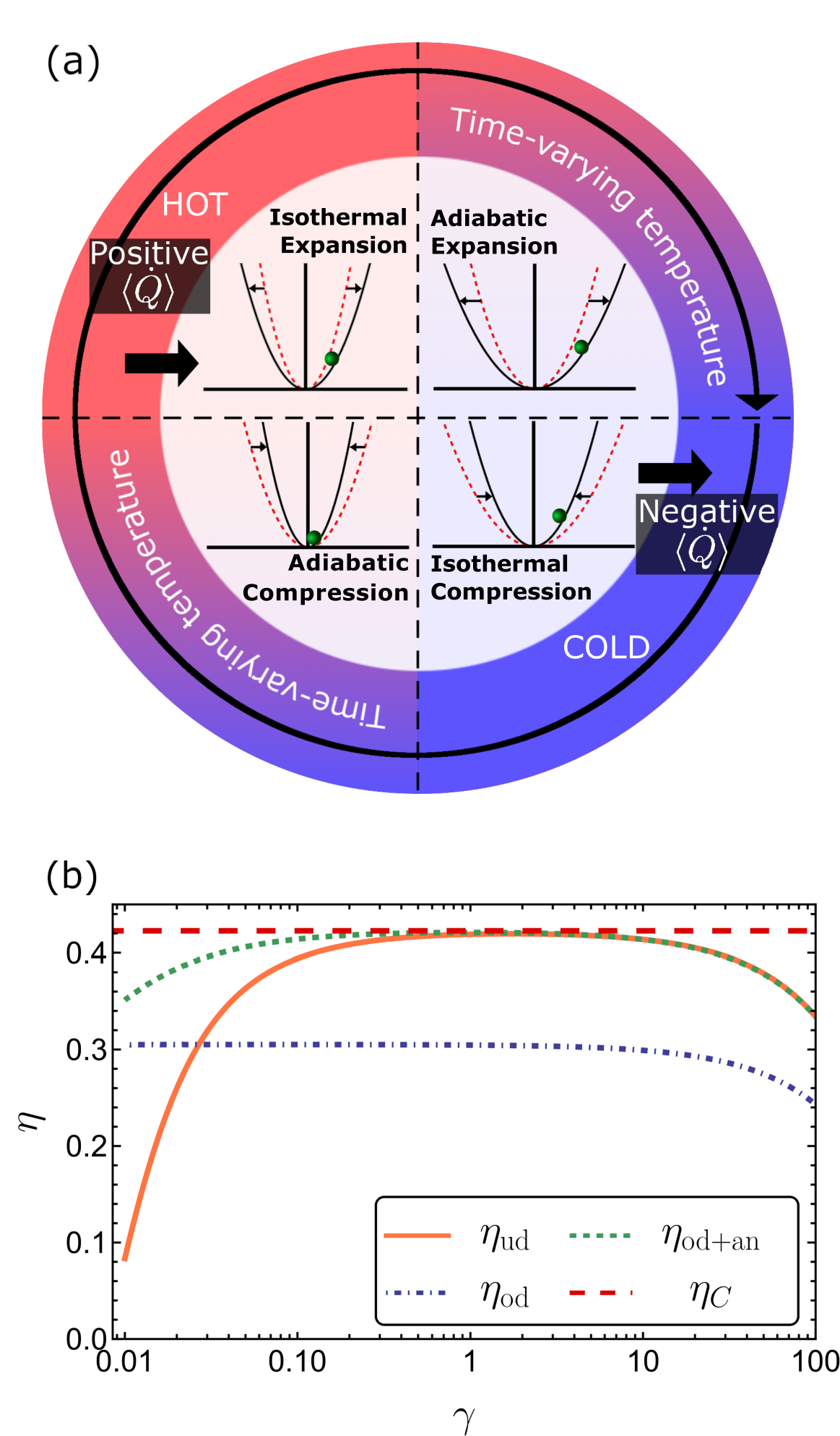}
	\caption{Numerical analysis of a finite-time Brownian Carnot engine. (a) Schematic diagram of a Brownian Carnot Engine depicting its four stages of operation in each time-cycle. The figure shows the thermodynamic cycle of the engine in a clockwise direction, consisting of four key processes: isothermal compression, adiabatic compression, isothermal expansion, and adiabatic expansion. 
    The temperature of the surrounding heat bath is represented by the outer colored ring, while the compression and expansion processes are indicated by variations in potential strength within each quadrant. (b) Efficiencies $\eta_{\rm ud}$, $\eta_{\rm od}$, and $\eta_{\rm od + an}$ as a function of $\gamma$ for the Brownian Carnot engine. The Carnot efficiency bound $\eta_C$ is also shown in the plot for reference.}
	\label{Fig_2}
    \end{figure}
	
	\subsection*{Efficiency of heat engine}\label{Sec:Results_Sub:Efficiency}
	The effects of the thermodynamic anomalies can be significant in engine systems. As an example, consider a Brownian Carnot engine experimentally realized in Ref.~\cite{Martinez2016}. The schematic diagram of the engine is depicted in Fig.~\hyperref[Fig_2]{2(a)}.  In this engine, a Brownian particle is confined in harmonic potential with a time-dependent stiffness $k(t)$: during the compression phase $0\leq t < t_{\rm p}/2$, $k(t)= k_0+k_1 t^2/t_{\rm p}^2$; during the expansion phase $t_{\rm p}/2\leq t < t_{\rm p}$, $k(t)= k_0+k_1(t_{\rm p} - t)^2/t_{\rm p}^2$, where $t_{\rm p}$ denotes the period of the engine. Each cycle of the engine is divided into four processes based on the temperature variation: (i) an isothermal process, $T(t)=T_{\rm c}$ for $0\leq t < t_{\rm p}/4$; (ii) an adiabatic process, $T(t)=T_{\rm c}\sqrt{k(t)/k(t_{\rm p}/4)}$ for $t_{\rm p}/4 \leq t < t_{\rm p}/2$; (iii) an isothermal process, $T(t)=T_{\rm h}$ for $t_{\rm p}/2 \leq t < 3t_{\rm p}/4$; and (iv) an adiabatic process, $T(t)=T_{\rm h}\sqrt{k(t)/k(3t_{\rm p}/4)}$ for $3t_{\rm p}/4 \leq t < t_{\rm p}$. The time variations of $k(t)$ and $T(t)$ are visually presented in Supplementary Section VI~\cite{supp_NC}. 
	
    Under this setup, we evaluate the three different efficiencies: $\eta_{\rm ud}$, $\eta_{\rm od}$, and $\eta_{\rm ud+an}$ using the underdamped formulation, overdamped formulation, and overdamped formulation incorporating the heat anomaly, respectively. The details are presented in the Methods section.
    Figure~\hyperref[Fig_2]{2(b)} shows the resulting plots of $\eta_{\rm ud}$, $\eta_{\rm od}$, and $\eta_{\rm od+an}$ as functions of $\gamma$. Since $m$ is set to $O(1)$ value, the overdamped approximation with $z=0$ is expected in the large-$\gamma$ regime. Thus, $\mathcal{A}_Q$ for $z=0$ is used to evaluate $\eta_{\rm od+an}$. Note that the contribution of the anomaly to heat becomes more significant in the large-$\gamma$ regime, as presented in Supplementary Section II~\cite{supp_NC}. This results in the efficiency discrepancy between $\eta_{\rm od}$ and $\eta_{\rm od+an}$. 
	As $\gamma$ increases, only $\eta_{\rm od+an}$ converges to $\eta_{\rm ud}$, whereas $\eta_{\rm od}$ deviates significantly from the other two. This highlights the substantial impact of the anomalies on the accurate calculation of thermodynamic quantities in heat engines.

	%
	%
	
	\subsection*{Estimation of kinetic energy}
	In some cases, evaluating kinetic energy in overdamped systems is necessary for accurately estimating thermodynamic quantities~\cite{Martinez2016, Roldan2014}. However, directly measuring kinetic energy in such systems is experimentally challenging, as velocity rapidly relaxes to equilibrium. Therefore, state-of-the-art experimental setups~\cite{Li2010, Huang2011, Kheifets2014} are required for precise kinetic energy measurements.
	To overcome this technical difficulty, a method, called time-averaged velocity (TAV) method, for estimating kinetic energy was proposed, where kinetic energy is inferred from measurements of the mean square velocity sampled at frequencies several orders of magnitude lower than the velocity relaxation frequency~\cite{Roldan2014}. However, the TAV method,  is limited to systems with a Brownian particle trapped in a harmonic potential and is accurate only for quasi-static processes, failing for moderately fast temperature variations, as shown in Supplementary Section VII~\cite{supp_NC}. In contrast, Eq.~\eqref{eq:kineticE} directly provides an accurate kinetic energy estimate even for rapidly varying temperatures. Performance comparisons between the TAV method and our approach are presented in Supplementary Section VII~\cite{supp_NC}.

	%
	%
	\section*{Discussion}\label{Sec:Discussion}
	We derive explicit expressions for heat and EP anomalies in systems immersed in an environment with a general time-varying temperature. 
	Using a cyclic engine as an example, we demonstrate that thermodynamic quantities such as heat and efficiency estimated within the overdamped description can significantly deviate from their correct values if these anomalies are not properly considered.
	An important point is that these anomalies depend on the exponent $z$, which relates the two scaling parameters, $\nu$ and $\tau$, introduced for the systematic overdamped approximation. Therefore, it is crucial to estimate $z$ for precise calculation of thermodynamic quantities. We propose an experimental method to estimate or control $z$ by adjusting the amplitudes of force and temperature.
	
	Our results allow for accurate estimation of thermodynamic quantities in overdamped systems without requiring complex or challenging experimental techniques to directly measure the system's fast-relaxing velocity. Instead, by simply incorporating the anomalies into the corresponding quantities computed within the overdamped description, we can achieve accurate measurements. Furthermore, our findings provide a straightforward method for estimating kinetic energy even for overdamped systems. Consequently, our results offer the systematic way for an accurate study of thermodynamics in a wide range of overdamped systems with time-varying temperature.

	\section*{Methods}\label{Sec:Method}
	
	\subsection*{Brinkman's hierarchy method with two-scaling parameters} \label{secA:Brinkman_two_scaling}
	
	Define $\bar{P}_{\rm ud} \equiv \psi_0^{-1}  P_{\rm ud} (x,v,t)$, where $\psi_0 = N e^{-\Phi/2}$ with the normalization factor $N \equiv \left[ 2 \pi T(t)/m\right]^{-1/4}$ and $\Phi \equiv \frac{mv^2}{2T(t)}$. Using the $n$th eigenfunction $\psi_n$ of a harmonic oscillator given by
	\begin{align} \label{eqA:psin-def}
		\psi_n  = \frac{1}{\sqrt{2^n n!} } \psi_0 H_n \left(\sqrt{\frac{m}{2T(t)}} v \right)\;,
	\end{align}
	where $H_n(x)$ denotes the Hermite polynomial, $\bar{P}_{\rm ud}$ can be expanded as
	\begin{equation} \label{eqA:Pbar-exp}
		\bar{P}_{\rm ud} = \sum_{n=0}^{\infty} c_n(x,t) \psi_n\;.
	\end{equation}
	Note that $c_0(x,t)=P_{\rm od}(x,t)$ as explained in Supplementary Section I~\cite{supp_NC}.
	Substituting Eq.~\eqref{eqA:Pbar-exp} into the Hermitianized Fokker-Planck equation yields the following coupled equation for the coefficients $c_n$ (for $n\geq 0$):
	\begin{widetext}
		\begin{align}
			\label{eqA:cn-eqn}
			\partial_t c_n = -\sqrt{\frac{(n+1)T(t)}{m}}\partial_x c_{n+1} + \frac{\sqrt{n}}{\sqrt{m T(t)}} \left[ f(x,\lambda_t) -T(t)\partial_x \right] c_{n-1} - \frac{n \gamma}{m}c_n - \frac{\dot{T}(t)}{2 T(t)} \left( \sqrt{n(n-1)} c_{n-2} +n c_n  \right)\;.
		\end{align}
	\end{widetext}
	The detailed derivation of Eq.~\eqref{eqA:cn-eqn} is provided in Supplementary Section I~\cite{supp_NC}. 
	
	Furthermore, it is shown in SM that the high-viscosity and the small-mass limits result in different expressions for heat and EP. To address this in a systematic way, here, we present a unified perturbative scheme, Brinkman's hierarchy with two-scaling parameters, capable of exploring not only these two limits but also the intermediate regimes.
	
	To achieve this, we convert Eq.~\eqref{eqA:cn-eqn} into a dimensionless form by introducing the characteristic time and length scales of the overdamped system, denoted as $\tau_\text{od}$ and $l_\text{od}$, respectively. 
	Using these, we define the dimensionless time, position, and $n$th coefficient as $\bar{t} \equiv t/\tau_\text{od}$, $\bar{x} \equiv x/l_\text{od}$, and $\bar{c}_n \equiv l_\text{od} c_n$, respectively. Additionally, we introduce a dimensionless temperature $\bar{T}(t) \equiv T(t)/T_0$, where $T_0$ represents the typical energy scale of the system.
	These definitions allow us to specify the typical velocities of the underdamped system, $V_\text{ud} \equiv \sqrt{T_0/m}$, and the overdamped system, $V_\text{od} \equiv l_\text{od}/\tau_\text{od}$. Using the quantities defined thus far, we can rewrite Eq.~\eqref{eqA:cn-eqn} in a dimensionless form as follows:
	\begin{widetext}
		\begin{align}
			\label{eqA:cbn-eqn_dimless}
			\partial_{\bar{t}} \bar{c}_n = -\nu \sqrt{(n+1)\bar{T}(t)}\partial_{\bar{x}} \bar{c}_{n+1} + \nu \sqrt{n\bar{T}(t)}\left[ \frac{\bar{f}(x,\lambda_t)}{\bar{T}(t)} -\partial_{\bar{x}} \right] \bar{c}_{n-1} - \tau n \bar{c}_n - \frac{\dot{\bar{T}}(t)}{2 \bar{T}(t)} \left( \sqrt{n(n-1)} \bar{c}_{n-2} +n \bar{c}_n  \right)~,
		\end{align}
	\end{widetext}
	where $\nu \equiv V_\text{ud}/V_\text{od}$, $\tau \equiv \tau_\text{od}/\tau_{\rm r}$, $\dot{\bar{T}}=d\bar{T}/d\bar{t}$, and $\bar{f} \equiv (l_\text{od}/T_0) f$ is the dimensionless force. 
	
	Instead of the two parameters $m$ and $\gamma$ in Eq.~\eqref{eqA:cn-eqn}, whose magnitudes govern the overdamped approximation, the two dimensionless parameters $\tau$ and $\nu$ in Eq.~\eqref{eqA:cbn-eqn_dimless} now play this role. Here, $\tau$ is large ($\tau \gg 1$) because we focus on the time scales $\tau_{\rm r} \ll \tau_{\rm obs} \leq \tau_{\rm od} \approx \tau_{\rm tmp}$. We also anticipate $\nu \gg 1$, as $V_\text{ud} \gg V_\text{od}$ is typically expected. 
	To systematically expand the orders in Eq.~\eqref{eqA:cbn-eqn_dimless}, we need to establish the magnitude relationship between $\tau$ and $\nu$. For simplicity, we set $\nu = \tau^z$, where $z \geq 0$, indicating that $\nu$ is of order $O(\tau^z)$.
	We note that $z=0$ leads to $m \sim T_0 \tau_{\rm od}^2 / l_{\rm od}^2$. Thus, $m$ is a $O(1)$ quantity in the dimensionless form, which implies that $\gamma$ must be large to satisfy the condition $\tau \gg 1$. Therefore, $z=0$ corresponds to the high-viscosity limit.
	On the other hand, $z=1/2$ gives $\gamma \sim T_0 \tau_{\rm od} / l_{\rm od}^2$. In this case, $\gamma$ is a $O(1)$ quantity in the dimensionless equation, which implies that $m$ must be small. Therefore, $z=1/2$ corresponds to the small-mass limit. 
	Finally, $0 < z < 1/2$ represents an intermediate regime between the high-viscosity and small-mass limits.
	
	For $0 \leq z \leq 1/2$, substituting $\nu = \tau^z$ into Eq.~\eqref{eqA:cbn-eqn_dimless} and collecting the leading-order terms result in (for $n=0,1$), 
	\begin{align}
		&\partial_{\bar{t}} \bar{c}_0 = - \tau^{z} \sqrt{\bar{T}}\partial_{\bar{x}} \bar{c}_1\;,\label{eqA:FPbar-eqn0}\\
		&\bar{c}_1 = \tau^{z-1} \sqrt{\bar{T}(t)}\left( \bar{f}/\bar{T} - \partial_{\bar{x}}\right)\bar{c}_0\;.\label{eqA:FPbar-eqn1}
	\end{align}
	Plugging Eq.~\eqref{eqA:FPbar-eqn1} into Eq.~\eqref{eqA:FPbar-eqn0} leads to the following dimensionless FP equation:
	\begin{equation}\label{eqA:FPbar-eqn}
		\partial_{\bar{t}} \bar{c}_0 = - \partial_{\bar{x}} \bar{J}(x,t)~.
	\end{equation}
	where the dimensionless probability current $\bar{J}(x,t)$ is given by
	\begin{equation}
		\bar{J}(x,t) = \tau^{2z-1} \left( \bar{f} - \bar{T} \partial_{\bar{x}} \right)\bar{c}_0~.
	\end{equation}
	As $\bar{J}(x,t)$ diverges for $z>1/2$, which is physically infeasible, we restrict our attention to the regime $0 \leq z \leq 1/2$. Note that if we convert the dimensionless variables in Eq.~\eqref{eqA:FPbar-eqn} back to their original forms, the equation becomes identical to Eq.~\eqref{FP-od}. This shows that the dynamic equations for overdamped systems are independent of $z$.

	\subsection*{Three different efficiencies: $\eta_{\rm ud}$, $\eta_{\rm od}$, and $\eta_{\rm od+an}$}
	
	Under the setup described in the ``Efficiency of heat engine'' section, we numerically evaluate the engine efficiency $\eta = \langle W\rangle/\langle Q_{\rm in}\rangle$ in a periodic steady state, where $\langle W\rangle = \int_{t=0}^{t_{\rm p}}dt\dot{k} \langle x^2 \rangle /2$ represents work done by the engine and $\langle Q_{\rm in}\rangle= \int_0^{t_{\rm p}}dt \Theta(\langle \dot{Q}\rangle) \langle \dot{Q} \rangle$ denotes the heat input to the engine. Here, the Heaviside function $\Theta(x)$ is defined as $\Theta(x)=1$ for $x>0$ and $\Theta(x)=0$ otherwise. For this calculation, the parameters are set as $k_0=2$, $k_1=64$, $T_c = 300$, $T_h =T_c\sqrt{k(t_{\rm p}/2)/k(t_{\rm p}/4)}$, $t_{\rm p}=10^3$, and $m=1$. 
	We evaluate the input heat in three different ways: using the underdamped formulation~\eqref{eq:Q_ud_1}, $\langle Q_{\rm in}\rangle_{\rm ud} = \int_0^{t_{\rm p}}dt \Theta(\langle \dot{Q}\rangle_{\rm ud}) \langle \dot{Q} \rangle_{\rm ud}$; using the overdamped formulation~\eqref{eq:heat_rate_over_conv}, $\langle Q_{\rm in}\rangle_{\rm od} = \int_0^{t_{\rm p}}dt \Theta(\langle \dot{Q}\rangle_{\rm od}) \langle \dot{Q} \rangle_{\rm od}$; and using the overdamped formulation with the addition of the heat anomaly~\eqref{dQ-ud}, $\langle Q_{\rm in}\rangle_{\rm od+an} =  \int_0^{t_{\rm p}}dt \Theta(\langle \dot{Q}\rangle_{\rm od} +\mathcal{A}_Q) (\langle \dot{Q} \rangle_{\rm od} +\mathcal{A}_Q)$. 
	Note that $\langle W \rangle$ does not depend on whether the underlying dynamics are underdamped or overdamped, as the work is evaluated solely using position trajectories. Then, the efficiency can also be defined in three ways: $\eta_{\rm ud} \equiv \langle W\rangle / \langle Q_{\rm in}\rangle_{\rm ud}$, $\eta_{\rm od} \equiv \langle W\rangle / \langle Q_{\rm in}\rangle_{\rm od}$, and $\eta_{\rm od+an} \equiv \langle W\rangle / \langle Q_{\rm in}\rangle_{\rm od+an}$.
	
	%
	%
	\section*{Data availability}
	The data that support the findings of this study are available from the corresponding
	author upon reasonable request.
	
	%
	%
	\section*{Code availability}
	Source code is available from the corresponding authors upon reasonable request.

	%
	%

\begin{thebibliography}{46}%
		\makeatletter
		\providecommand \@ifxundefined [1]{%
			\@ifx{#1\undefined}
		}%
		\providecommand \@ifnum [1]{%
			\ifnum #1\expandafter \@firstoftwo
			\else \expandafter \@secondoftwo
			\fi
		}%
		\providecommand \@ifx [1]{%
			\ifx #1\expandafter \@firstoftwo
			\else \expandafter \@secondoftwo
			\fi
		}%
		\providecommand \natexlab [1]{#1}%
		\providecommand \enquote  [1]{``#1''}%
		\providecommand \bibnamefont  [1]{#1}%
		\providecommand \bibfnamefont [1]{#1}%
		\providecommand \citenamefont [1]{#1}%
		\providecommand \href@noop [0]{\@secondoftwo}%
		\providecommand \href [0]{\begingroup \@sanitize@url \@href}%
		\providecommand \@href[1]{\@@startlink{#1}\@@href}%
		\providecommand \@@href[1]{\endgroup#1\@@endlink}%
		\providecommand \@sanitize@url [0]{\catcode `\\12\catcode `\$12\catcode
			`\&12\catcode `\#12\catcode `\^12\catcode `\_12\catcode `\%12\relax}%
		\providecommand \@@startlink[1]{}%
		\providecommand \@@endlink[0]{}%
		\providecommand \url  [0]{\begingroup\@sanitize@url \@url }%
		\providecommand \@url [1]{\endgroup\@href {#1}{\urlprefix }}%
		\providecommand \urlprefix  [0]{URL }%
		\providecommand \Eprint [0]{\href }%
		\providecommand \doibase [0]{https://doi.org/}%
		\providecommand \selectlanguage [0]{\@gobble}%
		\providecommand \bibinfo  [0]{\@secondoftwo}%
		\providecommand \bibfield  [0]{\@secondoftwo}%
		\providecommand \translation [1]{[#1]}%
		\providecommand \BibitemOpen [0]{}%
		\providecommand \bibitemStop [0]{}%
		\providecommand \bibitemNoStop [0]{.\EOS\space}%
		\providecommand \EOS [0]{\spacefactor3000\relax}%
		\providecommand \BibitemShut  [1]{\csname bibitem#1\endcsname}%
		\let\auto@bib@innerbib\@empty
        \bibitem [{\citenamefont {Whalen}\ \emph {et~al.}(2003)\citenamefont {Whalen},
			\citenamefont {Thompson}, \citenamefont {Bahr}, \citenamefont {Richards},\
			and\ \citenamefont {Richards}}]{Whalen2003}%
		\BibitemOpen
		\bibfield  {author} {\bibinfo {author} {\bibfnamefont {S.}~\bibnamefont
				{Whalen}}, \bibinfo {author} {\bibfnamefont {M.}~\bibnamefont {Thompson}},
			\bibinfo {author} {\bibfnamefont {D.}~\bibnamefont {Bahr}}, \bibinfo {author}
			{\bibfnamefont {C.}~\bibnamefont {Richards}},\ and\ \bibinfo {author}
			{\bibfnamefont {R.}~\bibnamefont {Richards}},\ }\bibfield  {title} {\bibinfo
			{title} {Design, fabrication and testing of the p3 micro heat engine},\
		}\href {https://doi.org/10.31438/trf.hh2002.6} {\bibfield  {journal}
			{\bibinfo  {journal} {Sensors and Actuators A: Physical}\ }\textbf {\bibinfo
				{volume} {104}},\ \bibinfo {pages} {290} (\bibinfo {year}
			{2003})}\BibitemShut {NoStop}%
        \bibitem [{\citenamefont {Blickle}\ and\ \citenamefont
			{Bechinger}(2012)}]{Blickle2012}%
		\BibitemOpen
		\bibfield  {author} {\bibinfo {author} {\bibfnamefont {V.}~\bibnamefont
				{Blickle}}\ and\ \bibinfo {author} {\bibfnamefont {C.}~\bibnamefont
				{Bechinger}},\ }\bibfield  {title} {\bibinfo {title} {Realization of a
				micrometre-sized stochastic heat engine},\ }\href
		{https://doi.org/10.1038/nphys2163} {\bibfield  {journal} {\bibinfo
				{journal} {Nature Physics}\ }\textbf {\bibinfo {volume} {8}},\ \bibinfo
			{pages} {143} (\bibinfo {year} {2012})}\BibitemShut {NoStop}%
		\bibitem [{\citenamefont {Steeneken}\ \emph {et~al.}(2011)\citenamefont
			{Steeneken}, \citenamefont {Le~Phan}, \citenamefont {Goossens}, \citenamefont
			{Koops}, \citenamefont {Brom}, \citenamefont {Van~der Avoort},\ and\
			\citenamefont {Van~Beek}}]{Steeneken2011}%
		\BibitemOpen
		\bibfield  {author} {\bibinfo {author} {\bibfnamefont {P.}~\bibnamefont
				{Steeneken}}, \bibinfo {author} {\bibfnamefont {K.}~\bibnamefont {Le~Phan}},
			\bibinfo {author} {\bibfnamefont {M.}~\bibnamefont {Goossens}}, \bibinfo
			{author} {\bibfnamefont {G.}~\bibnamefont {Koops}}, \bibinfo {author}
			{\bibfnamefont {G.}~\bibnamefont {Brom}}, \bibinfo {author} {\bibfnamefont
				{C.}~\bibnamefont {Van~der Avoort}},\ and\ \bibinfo {author} {\bibfnamefont
				{J.}~\bibnamefont {Van~Beek}},\ }\bibfield  {title} {\bibinfo {title}
			{Piezoresistive heat engine and refrigerator},\ }\href
		{https://doi.org/10.1038/nphys1871} {\bibfield  {journal} {\bibinfo
				{journal} {Nature Physics}\ }\textbf {\bibinfo {volume} {7}},\ \bibinfo
			{pages} {354} (\bibinfo {year} {2011})}\BibitemShut {NoStop}%
        \bibitem [{\citenamefont {Brantut}\ \emph {et~al.}(2013)\citenamefont
			{Brantut}, \citenamefont {Grenier}, \citenamefont {Meineke}, \citenamefont
			{Stadler}, \citenamefont {Krinner}, \citenamefont {Kollath}, \citenamefont
			{Esslinger},\ and\ \citenamefont {Georges}}]{Brantut2013}%
		\BibitemOpen
		\bibfield  {author} {\bibinfo {author} {\bibfnamefont {J.-P.}\ \bibnamefont
				{Brantut}}, \bibinfo {author} {\bibfnamefont {C.}~\bibnamefont {Grenier}},
			\bibinfo {author} {\bibfnamefont {J.}~\bibnamefont {Meineke}}, \bibinfo
			{author} {\bibfnamefont {D.}~\bibnamefont {Stadler}}, \bibinfo {author}
			{\bibfnamefont {S.}~\bibnamefont {Krinner}}, \bibinfo {author} {\bibfnamefont
				{C.}~\bibnamefont {Kollath}}, \bibinfo {author} {\bibfnamefont
				{T.}~\bibnamefont {Esslinger}},\ and\ \bibinfo {author} {\bibfnamefont
				{A.}~\bibnamefont {Georges}},\ }\bibfield  {title} {\bibinfo {title} {A
				thermoelectric heat engine with ultracold atoms},\ }\href
		{https://doi.org/10.1126/science.1242308} {\bibfield  {journal} {\bibinfo
				{journal} {Science}\ }\textbf {\bibinfo {volume} {342}},\ \bibinfo {pages}
			{713} (\bibinfo {year} {2013})}\BibitemShut {NoStop}%
        \bibitem [{\citenamefont {Argun}\ \emph {et~al.}(2017)\citenamefont {Argun},
			\citenamefont {Soni}, \citenamefont {Dabelow}, \citenamefont {Bo},
			\citenamefont {Pesce}, \citenamefont {Eichhorn},\ and\ \citenamefont
			{Volpe}}]{Argun2017}%
		\BibitemOpen
		\bibfield  {author} {\bibinfo {author} {\bibfnamefont {A.}~\bibnamefont
				{Argun}}, \bibinfo {author} {\bibfnamefont {J.}~\bibnamefont {Soni}},
			\bibinfo {author} {\bibfnamefont {L.}~\bibnamefont {Dabelow}}, \bibinfo
			{author} {\bibfnamefont {S.}~\bibnamefont {Bo}}, \bibinfo {author}
			{\bibfnamefont {G.}~\bibnamefont {Pesce}}, \bibinfo {author} {\bibfnamefont
				{R.}~\bibnamefont {Eichhorn}},\ and\ \bibinfo {author} {\bibfnamefont
				{G.}~\bibnamefont {Volpe}},\ }\bibfield  {title} {\bibinfo {title}
			{Experimental realization of a minimal microscopic heat engine},\ }\href
		{https://doi.org/10.1103/physreve.96.052106} {\bibfield  {journal} {\bibinfo
				{journal} {Physical Review E}\ }\textbf {\bibinfo {volume} {96}},\ \bibinfo
			{pages} {052106} (\bibinfo {year} {2017})}\BibitemShut {NoStop}%
        \bibitem [{\citenamefont {Mart{\'\i}nez}\ \emph {et~al.}(2016)\citenamefont
			{Mart{\'\i}nez}, \citenamefont {Rold{\'a}n}, \citenamefont {Dinis},
			\citenamefont {Petrov}, \citenamefont {Parrondo},\ and\ \citenamefont
			{Rica}}]{Martinez2016}%
		\BibitemOpen
		\bibfield  {author} {\bibinfo {author} {\bibfnamefont {I.~A.}\ \bibnamefont
				{Mart{\'\i}nez}}, \bibinfo {author} {\bibfnamefont {{\'E}.}~\bibnamefont
				{Rold{\'a}n}}, \bibinfo {author} {\bibfnamefont {L.}~\bibnamefont {Dinis}},
			\bibinfo {author} {\bibfnamefont {D.}~\bibnamefont {Petrov}}, \bibinfo
			{author} {\bibfnamefont {J.~M.}\ \bibnamefont {Parrondo}},\ and\ \bibinfo
			{author} {\bibfnamefont {R.~A.}\ \bibnamefont {Rica}},\ }\bibfield  {title}
		{\bibinfo {title} {Brownian carnot engine},\ }\href
		{https://doi.org/10.1038/nphys3518} {\bibfield  {journal} {\bibinfo
				{journal} {Nature physics}\ }\textbf {\bibinfo {volume} {12}},\ \bibinfo
			{pages} {67} (\bibinfo {year} {2016})}\BibitemShut {NoStop}%
		\bibitem [{\citenamefont {Krishnamurthy}\ \emph {et~al.}(2016)\citenamefont
			{Krishnamurthy}, \citenamefont {Ghosh}, \citenamefont {Chatterji},
			\citenamefont {Ganapathy},\ and\ \citenamefont {Sood}}]{Krishnamurthy2016}%
		\BibitemOpen
		\bibfield  {author} {\bibinfo {author} {\bibfnamefont {S.}~\bibnamefont
				{Krishnamurthy}}, \bibinfo {author} {\bibfnamefont {S.}~\bibnamefont
				{Ghosh}}, \bibinfo {author} {\bibfnamefont {D.}~\bibnamefont {Chatterji}},
			\bibinfo {author} {\bibfnamefont {R.}~\bibnamefont {Ganapathy}},\ and\
			\bibinfo {author} {\bibfnamefont {A.}~\bibnamefont {Sood}},\ }\bibfield
		{title} {\bibinfo {title} {A micrometre-sized heat engine operating between
				bacterial reservoirs},\ }\href {https://doi.org/10.1038/nphys3870} {\bibfield
			{journal} {\bibinfo  {journal} {Nature Physics}\ }\textbf {\bibinfo {volume}
				{12}},\ \bibinfo {pages} {1134} (\bibinfo {year} {2016})}\BibitemShut
		{NoStop}%
		\bibitem [{\citenamefont {Krishnamurthy}\ \emph {et~al.}(2023)\citenamefont
			{Krishnamurthy}, \citenamefont {Ganapathy},\ and\ \citenamefont
			{Sood}}]{Krishnamurthy2023}%
		\BibitemOpen
		\bibfield  {author} {\bibinfo {author} {\bibfnamefont {S.}~\bibnamefont
				{Krishnamurthy}}, \bibinfo {author} {\bibfnamefont {R.}~\bibnamefont
				{Ganapathy}},\ and\ \bibinfo {author} {\bibfnamefont {A.}~\bibnamefont
				{Sood}},\ }\bibfield  {title} {\bibinfo {title} {Overcoming power-efficiency
				tradeoff in a micro heat engine by engineered system-bath interactions},\
		}\href {https://doi.org/10.1038/s41467-023-42350-y} {\bibfield  {journal}
			{\bibinfo  {journal} {Nature Communications}\ }\textbf {\bibinfo {volume}
				{14}},\ \bibinfo {pages} {6842} (\bibinfo {year} {2023})}\BibitemShut
		{NoStop}%
        \bibitem [{\citenamefont {Widder}\ and\ \citenamefont
			{Titulaer}(1989)}]{Widder1989}%
		\BibitemOpen
		\bibfield  {author} {\bibinfo {author} {\bibfnamefont {M.}~\bibnamefont
				{Widder}}\ and\ \bibinfo {author} {\bibfnamefont {U.}~\bibnamefont
				{Titulaer}},\ }\bibfield  {title} {\bibinfo {title} {Brownian motion in a
				medium with inhomogeneous temperature},\ }\href
		{https://doi.org/https://doi.org/10.1016/0378-4371(89)90259-8} {\bibfield
			{journal} {\bibinfo  {journal} {Physica A: Statistical Mechanics and its
					Applications}\ }\textbf {\bibinfo {volume} {154}},\ \bibinfo {pages} {452}
			(\bibinfo {year} {1989})}\BibitemShut {NoStop}%
		\bibitem [{\citenamefont {Hondou}\ and\ \citenamefont
			{Sekimoto}(2000)}]{Hondou2000}%
		\BibitemOpen
		\bibfield  {author} {\bibinfo {author} {\bibfnamefont {T.}~\bibnamefont
				{Hondou}}\ and\ \bibinfo {author} {\bibfnamefont {K.}~\bibnamefont
				{Sekimoto}},\ }\bibfield  {title} {\bibinfo {title} {Unattainability of
				carnot efficiency in the brownian heat engine},\ }\href
		{https://doi.org/https://doi.org/10.1103/physreve.62.6021} {\bibfield
			{journal} {\bibinfo  {journal} {Physical Review E}\ }\textbf {\bibinfo
				{volume} {62}},\ \bibinfo {pages} {6021} (\bibinfo {year}
			{2000})}\BibitemShut {NoStop}%
		\bibitem [{\citenamefont {Celani}\ \emph {et~al.}(2012)\citenamefont {Celani},
			\citenamefont {Bo}, \citenamefont {Eichhorn},\ and\ \citenamefont
			{Aurell}}]{Celani2012}%
		\BibitemOpen
		\bibfield  {author} {\bibinfo {author} {\bibfnamefont {A.}~\bibnamefont
				{Celani}}, \bibinfo {author} {\bibfnamefont {S.}~\bibnamefont {Bo}}, \bibinfo
			{author} {\bibfnamefont {R.}~\bibnamefont {Eichhorn}},\ and\ \bibinfo
			{author} {\bibfnamefont {E.}~\bibnamefont {Aurell}},\ }\bibfield  {title}
		{\bibinfo {title} {Anomalous thermodynamics at the microscale},\ }\href
		{https://doi.org/10.1103/physrevlett.109.260603} {\bibfield  {journal}
			{\bibinfo  {journal} {Physical review letters}\ }\textbf {\bibinfo {volume}
				{109}},\ \bibinfo {pages} {260603} (\bibinfo {year} {2012})}\BibitemShut
		{NoStop}%
		\bibitem [{\citenamefont {Bo}\ \emph {et~al.}(2013)\citenamefont {Bo},
			\citenamefont {Aurell}, \citenamefont {Eichhorn},\ and\ \citenamefont
			{Celani}}]{Bo2013}%
		\BibitemOpen
		\bibfield  {author} {\bibinfo {author} {\bibfnamefont {S.}~\bibnamefont
				{Bo}}, \bibinfo {author} {\bibfnamefont {E.}~\bibnamefont {Aurell}}, \bibinfo
			{author} {\bibfnamefont {R.}~\bibnamefont {Eichhorn}},\ and\ \bibinfo
			{author} {\bibfnamefont {A.}~\bibnamefont {Celani}},\ }\bibfield  {title}
		{\bibinfo {title} {Optimal stochastic transport in inhomogeneous thermal
				environments},\ }\href
		{https://doi.org/https://doi.org/10.1209/0295-5075/103/10010} {\bibfield
			{journal} {\bibinfo  {journal} {Europhysics letters}\ }\textbf {\bibinfo
				{volume} {103}},\ \bibinfo {pages} {10010} (\bibinfo {year}
			{2013})}\BibitemShut {NoStop}%
		\bibitem [{\citenamefont {Kawaguchi}\ and\ \citenamefont
			{Nakayama}(2013)}]{Kawaguchi2013}%
		\BibitemOpen
		\bibfield  {author} {\bibinfo {author} {\bibfnamefont {K.}~\bibnamefont
				{Kawaguchi}}\ and\ \bibinfo {author} {\bibfnamefont {Y.}~\bibnamefont
				{Nakayama}},\ }\bibfield  {title} {\bibinfo {title} {Fluctuation theorem for
				hidden entropy production},\ }\href
		{https://doi.org/https://doi.org/10.1103/physreve.88.022147} {\bibfield
			{journal} {\bibinfo  {journal} {Physical Review E}\ }\textbf {\bibinfo
				{volume} {88}},\ \bibinfo {pages} {022147} (\bibinfo {year}
			{2013})}\BibitemShut {NoStop}%
		\bibitem [{\citenamefont {Polettini}(2013)}]{Polettini2013}%
		\BibitemOpen
		\bibfield  {author} {\bibinfo {author} {\bibfnamefont {M.}~\bibnamefont
				{Polettini}},\ }\bibfield  {title} {\bibinfo {title} {Diffusion in nonuniform
				temperature and its geometric analog},\ }\href
		{https://doi.org/https://doi.org/10.1103/physreve.87.032126} {\bibfield
			{journal} {\bibinfo  {journal} {Physical Review E - Statistical, Nonlinear,
					and Soft Matter Physics}\ }\textbf {\bibinfo {volume} {87}},\ \bibinfo
			{pages} {032126} (\bibinfo {year} {2013})}\BibitemShut {NoStop}%
		\bibitem [{\citenamefont {Sancho}(2015)}]{Sancho2015}%
		\BibitemOpen
		\bibfield  {author} {\bibinfo {author} {\bibfnamefont {J.}~\bibnamefont
				{Sancho}},\ }\bibfield  {title} {\bibinfo {title} {Brownian colloids in
				underdamped and overdamped regimes with nonhomogeneous temperature},\ }\href
		{https://doi.org/https://doi.org/10.1103/physreve.92.062110} {\bibfield
			{journal} {\bibinfo  {journal} {Physical Review E}\ }\textbf {\bibinfo
				{volume} {92}},\ \bibinfo {pages} {062110} (\bibinfo {year}
			{2015})}\BibitemShut {NoStop}%
		\bibitem [{\citenamefont {Marino}\ \emph {et~al.}(2016)\citenamefont {Marino},
			\citenamefont {Eichhorn},\ and\ \citenamefont {Aurell}}]{Marino2016}%
		\BibitemOpen
		\bibfield  {author} {\bibinfo {author} {\bibfnamefont {R.}~\bibnamefont
				{Marino}}, \bibinfo {author} {\bibfnamefont {R.}~\bibnamefont {Eichhorn}},\
			and\ \bibinfo {author} {\bibfnamefont {E.}~\bibnamefont {Aurell}},\
		}\bibfield  {title} {\bibinfo {title} {Entropy production of a brownian
				ellipsoid in the overdamped limit},\ }\href
		{https://doi.org/https://doi.org/10.1103/physreve.93.012132} {\bibfield
			{journal} {\bibinfo  {journal} {Physical Review E}\ }\textbf {\bibinfo
				{volume} {93}},\ \bibinfo {pages} {012132} (\bibinfo {year}
			{2016})}\BibitemShut {NoStop}%
\bibitem [{\citenamefont {Schmiedl}\ and\ \citenamefont
			{Seifert}(2007)}]{Schmiedl2007_2}%
		\BibitemOpen
		\bibfield  {author} {\bibinfo {author} {\bibfnamefont {T.}~\bibnamefont
				{Schmiedl}}\ and\ \bibinfo {author} {\bibfnamefont {U.}~\bibnamefont
				{Seifert}},\ }\bibfield  {title} {\bibinfo {title} {Efficiency at maximum
				power: An analytically solvable model for stochastic heat engines},\ }\href
		{https://doi.org/10.1209/0295-5075/81/20003} {\bibfield  {journal} {\bibinfo
				{journal} {Europhysics letters}\ }\textbf {\bibinfo {volume} {81}},\ \bibinfo
			{pages} {20003} (\bibinfo {year} {2007})}\BibitemShut {NoStop}%
		\bibitem [{\citenamefont {Arold}\ \emph {et~al.}(2018)\citenamefont {Arold},
			\citenamefont {Dechant},\ and\ \citenamefont {Lutz}}]{Arold2019}%
		\BibitemOpen
		\bibfield  {author} {\bibinfo {author} {\bibfnamefont {D.}~\bibnamefont
				{Arold}}, \bibinfo {author} {\bibfnamefont {A.}~\bibnamefont {Dechant}},\
			and\ \bibinfo {author} {\bibfnamefont {E.}~\bibnamefont {Lutz}},\ }\bibfield
		{title} {\bibinfo {title} {Heat leakage in overdamped harmonic systems},\
		}\href {https://doi.org/10.1103/PhysRevE.97.022131} {\bibfield  {journal}
			{\bibinfo  {journal} {Phys. Rev. E}\ }\textbf {\bibinfo {volume} {97}},\
			\bibinfo {pages} {022131} (\bibinfo {year} {2018})}\BibitemShut {NoStop}%
		\bibitem [{\citenamefont {Awasthi}\ and\ \citenamefont
			{Dutta}(2022)}]{Awasthi2022}%
		\BibitemOpen
		\bibfield  {author} {\bibinfo {author} {\bibfnamefont {S.}~\bibnamefont
				{Awasthi}}\ and\ \bibinfo {author} {\bibfnamefont {S.~B.}\ \bibnamefont
				{Dutta}},\ }\bibfield  {title} {\bibinfo {title} {Oscillating states of
				driven langevin systems under large viscous drives},\ }\href
		{https://doi.org/10.1103/PhysRevE.106.064116} {\bibfield  {journal} {\bibinfo
				{journal} {Phys. Rev. E}\ }\textbf {\bibinfo {volume} {106}},\ \bibinfo
			{pages} {064116} (\bibinfo {year} {2022})}\BibitemShut {NoStop}%
		\bibitem [{\citenamefont {Parrondo}\ and\ \citenamefont
			{Espa{\~n}ol}(1996)}]{Parrondo1996}%
		\BibitemOpen
		\bibfield  {author} {\bibinfo {author} {\bibfnamefont {J.~M.}\ \bibnamefont
				{Parrondo}}\ and\ \bibinfo {author} {\bibfnamefont {P.}~\bibnamefont
				{Espa{\~n}ol}},\ }\bibfield  {title} {\bibinfo {title} {Criticism of
				feynman's analysis of the ratchet as an engine},\ }\href
		{https://doi.org/https://doi.org/10.1119/1.18393} {\bibfield  {journal}
			{\bibinfo  {journal} {American Journal of Physics}\ }\textbf {\bibinfo
				{volume} {64}},\ \bibinfo {pages} {1125} (\bibinfo {year}
			{1996})}\BibitemShut {NoStop}%
		\bibitem [{\citenamefont {Der{\'e}nyi}\ and\ \citenamefont
			{Astumian}(1999)}]{Derenyi1999}%
		\BibitemOpen
		\bibfield  {author} {\bibinfo {author} {\bibfnamefont {I.}~\bibnamefont
				{Der{\'e}nyi}}\ and\ \bibinfo {author} {\bibfnamefont {R.~D.}\ \bibnamefont
				{Astumian}},\ }\bibfield  {title} {\bibinfo {title} {Efficiency of brownian
				heat engines},\ }\href
		{https://doi.org/https://doi.org/10.1103/physreve.59.r6219} {\bibfield
			{journal} {\bibinfo  {journal} {Physical Review E}\ }\textbf {\bibinfo
				{volume} {59}},\ \bibinfo {pages} {R6219} (\bibinfo {year}
			{1999})}\BibitemShut {NoStop}%
		\bibitem [{\citenamefont {Van~den Broeck}(2005)}]{Van2005}%
		\BibitemOpen
		\bibfield  {author} {\bibinfo {author} {\bibfnamefont {C.}~\bibnamefont
				{Van~den Broeck}},\ }\bibfield  {title} {\bibinfo {title} {Thermodynamic
				efficiency at maximum power},\ }\href
		{https://doi.org/https://doi.org/10.1103/physrevlett.95.190602} {\bibfield
			{journal} {\bibinfo  {journal} {Physical review letters}\ }\textbf {\bibinfo
				{volume} {95}},\ \bibinfo {pages} {190602} (\bibinfo {year}
			{2005})}\BibitemShut {NoStop}%
		\bibitem [{\citenamefont {Rana}\ \emph {et~al.}(2014)\citenamefont {Rana},
			\citenamefont {Pal}, \citenamefont {Saha},\ and\ \citenamefont
			{Jayannavar}}]{Rana2014}%
		\BibitemOpen
		\bibfield  {author} {\bibinfo {author} {\bibfnamefont {S.}~\bibnamefont
				{Rana}}, \bibinfo {author} {\bibfnamefont {P.}~\bibnamefont {Pal}}, \bibinfo
			{author} {\bibfnamefont {A.}~\bibnamefont {Saha}},\ and\ \bibinfo {author}
			{\bibfnamefont {A.}~\bibnamefont {Jayannavar}},\ }\bibfield  {title}
		{\bibinfo {title} {Single-particle stochastic heat engine},\ }\href
		{https://doi.org/https://doi.org/10.1103/physreve.90.042146} {\bibfield
			{journal} {\bibinfo  {journal} {Physical review E}\ }\textbf {\bibinfo
				{volume} {90}},\ \bibinfo {pages} {042146} (\bibinfo {year}
			{2014})}\BibitemShut {NoStop}%
		\bibitem [{\citenamefont {Park}\ \emph {et~al.}(2016)\citenamefont {Park},
			\citenamefont {Lee},\ and\ \citenamefont {Noh}}]{Park2016}%
		\BibitemOpen
		\bibfield  {author} {\bibinfo {author} {\bibfnamefont {J.-M.}\ \bibnamefont
				{Park}}, \bibinfo {author} {\bibfnamefont {J.~S.}\ \bibnamefont {Lee}},\ and\
			\bibinfo {author} {\bibfnamefont {J.~D.}\ \bibnamefont {Noh}},\ }\bibfield
		{title} {\bibinfo {title} {Optimal tuning of a confined brownian information
				engine},\ }\href {https://doi.org/https://doi.org/10.1103/physreve.93.032146}
		{\bibfield  {journal} {\bibinfo  {journal} {Physical Review E}\ }\textbf
			{\bibinfo {volume} {93}},\ \bibinfo {pages} {032146} (\bibinfo {year}
			{2016})}\BibitemShut {NoStop}%
		\bibitem [{\citenamefont {Ro{\ss}nagel}\ \emph {et~al.}(2016)\citenamefont
			{Ro{\ss}nagel}, \citenamefont {Dawkins}, \citenamefont {Tolazzi},
			\citenamefont {Abah}, \citenamefont {Lutz}, \citenamefont {Schmidt-Kaler},\
			and\ \citenamefont {Singer}}]{Rossnagel2016}%
		\BibitemOpen
		\bibfield  {author} {\bibinfo {author} {\bibfnamefont {J.}~\bibnamefont
				{Ro{\ss}nagel}}, \bibinfo {author} {\bibfnamefont {S.~T.}\ \bibnamefont
				{Dawkins}}, \bibinfo {author} {\bibfnamefont {K.~N.}\ \bibnamefont
				{Tolazzi}}, \bibinfo {author} {\bibfnamefont {O.}~\bibnamefont {Abah}},
			\bibinfo {author} {\bibfnamefont {E.}~\bibnamefont {Lutz}}, \bibinfo {author}
			{\bibfnamefont {F.}~\bibnamefont {Schmidt-Kaler}},\ and\ \bibinfo {author}
			{\bibfnamefont {K.}~\bibnamefont {Singer}},\ }\bibfield  {title} {\bibinfo
			{title} {A single-atom heat engine},\ }\href
		{https://doi.org/10.1126/science.aad6320} {\bibfield  {journal} {\bibinfo
				{journal} {Science}\ }\textbf {\bibinfo {volume} {352}},\ \bibinfo {pages}
			{325} (\bibinfo {year} {2016})}\BibitemShut {NoStop}%
		\bibitem [{\citenamefont {Mart{\'\i}nez}\ \emph {et~al.}(2017)\citenamefont
			{Mart{\'\i}nez}, \citenamefont {Rold{\'a}n}, \citenamefont {Dinis},\ and\
			\citenamefont {Rica}}]{Martinez2017}%
		\BibitemOpen
		\bibfield  {author} {\bibinfo {author} {\bibfnamefont {I.~A.}\ \bibnamefont
				{Mart{\'\i}nez}}, \bibinfo {author} {\bibfnamefont {{\'E}.}~\bibnamefont
				{Rold{\'a}n}}, \bibinfo {author} {\bibfnamefont {L.}~\bibnamefont {Dinis}},\
			and\ \bibinfo {author} {\bibfnamefont {R.~A.}\ \bibnamefont {Rica}},\
		}\bibfield  {title} {\bibinfo {title} {Colloidal heat engines: a review},\
		}\href {https://doi.org/10.1039/c6sm00923a} {\bibfield  {journal} {\bibinfo
				{journal} {Soft matter}\ }\textbf {\bibinfo {volume} {13}} (\bibinfo {year}
			{2017})}\BibitemShut {NoStop}%
		\bibitem [{\citenamefont {Lee}\ \emph {et~al.}(2019)\citenamefont {Lee},
			\citenamefont {Park},\ and\ \citenamefont {Park}}]{Lee2019}%
		\BibitemOpen
		\bibfield  {author} {\bibinfo {author} {\bibfnamefont {J.~S.}\ \bibnamefont
				{Lee}}, \bibinfo {author} {\bibfnamefont {J.-M.}\ \bibnamefont {Park}},\ and\
			\bibinfo {author} {\bibfnamefont {H.}~\bibnamefont {Park}},\ }\bibfield
		{title} {\bibinfo {title} {Thermodynamic uncertainty relation for underdamped
				langevin systems driven by a velocity-dependent force},\ }\href
		{https://doi.org/10.1103/PhysRevE.100.062132} {\bibfield  {journal} {\bibinfo
				{journal} {Phys. Rev. E}\ }\textbf {\bibinfo {volume} {100}},\ \bibinfo
			{pages} {062132} (\bibinfo {year} {2019})}\BibitemShut {NoStop}%
		\bibitem [{\citenamefont {Holubec}\ and\ \citenamefont
			{Marathe}(2020)}]{Holubec2020_2}%
		\BibitemOpen
		\bibfield  {author} {\bibinfo {author} {\bibfnamefont {V.}~\bibnamefont
				{Holubec}}\ and\ \bibinfo {author} {\bibfnamefont {R.}~\bibnamefont
				{Marathe}},\ }\bibfield  {title} {\bibinfo {title} {Underdamped active
				brownian heat engine},\ }\href
		{https://doi.org/https://doi.org/10.1103/physreve.102.060101} {\bibfield
			{journal} {\bibinfo  {journal} {Physical Review E}\ }\textbf {\bibinfo
				{volume} {102}},\ \bibinfo {pages} {060101} (\bibinfo {year}
			{2020})}\BibitemShut {NoStop}%
		\bibitem [{\citenamefont {Chen}\ \emph {et~al.}(2022)\citenamefont {Chen},
			\citenamefont {Qi}, \citenamefont {Ge},\ and\ \citenamefont
			{Feng}}]{Chen2022}%
		\BibitemOpen
		\bibfield  {author} {\bibinfo {author} {\bibfnamefont {L.}~\bibnamefont
				{Chen}}, \bibinfo {author} {\bibfnamefont {C.}~\bibnamefont {Qi}}, \bibinfo
			{author} {\bibfnamefont {Y.}~\bibnamefont {Ge}},\ and\ \bibinfo {author}
			{\bibfnamefont {H.}~\bibnamefont {Feng}},\ }\bibfield  {title} {\bibinfo
			{title} {Thermal brownian heat engine with external and internal
				irreversibilities},\ }\href
		{https://doi.org/https://doi.org/10.1016/j.energy.2022.124582} {\bibfield
			{journal} {\bibinfo  {journal} {Energy}\ }\textbf {\bibinfo {volume} {255}},\
			\bibinfo {pages} {124582} (\bibinfo {year} {2022})}\BibitemShut {NoStop}%
		\bibitem [{\citenamefont {Majumdar}\ \emph {et~al.}(2022)\citenamefont
			{Majumdar}, \citenamefont {Saha},\ and\ \citenamefont
			{Marathe}}]{Majumdar2022}%
		\BibitemOpen
		\bibfield  {author} {\bibinfo {author} {\bibfnamefont {R.}~\bibnamefont
				{Majumdar}}, \bibinfo {author} {\bibfnamefont {A.}~\bibnamefont {Saha}},\
			and\ \bibinfo {author} {\bibfnamefont {R.}~\bibnamefont {Marathe}},\
		}\bibfield  {title} {\bibinfo {title} {Exactly solvable model of a passive
				brownian heat engine and its comparison with active engines},\ }\href
		{https://doi.org/https://doi.org/10.1088/1742-5468/ac7e3d} {\bibfield
			{journal} {\bibinfo  {journal} {Journal of Statistical Mechanics: Theory and
					Experiment}\ }\textbf {\bibinfo {volume} {2022}},\ \bibinfo {pages} {073206}
			(\bibinfo {year} {2022})}\BibitemShut {NoStop}%
		\bibitem [{\citenamefont {Risken}(1996)}]{Risken1996}%
		\BibitemOpen
		\bibfield  {author} {\bibinfo {author} {\bibfnamefont {H.}~\bibnamefont
				{Risken}},\ }\href {https://doi.org/10.1007/978-3-642-61544-3} {\emph
			{\bibinfo {title} {The Fokker-Planck Equation}}}\ (\bibinfo  {publisher}
		{Springer-Verlag Berlin Heidelberg},\ \bibinfo {year} {1996})\BibitemShut
		{NoStop}%
		\bibitem [{\citenamefont {Sekimoto}(1998)}]{Sekimoto1998}%
		\BibitemOpen
		\bibfield  {author} {\bibinfo {author} {\bibfnamefont {K.}~\bibnamefont
				{Sekimoto}},\ }\bibfield  {title} {\bibinfo {title} {{Langevin Equation and
					Thermodynamics}},\ }\href {https://doi.org/10.1143/PTPS.130.17} {\bibfield
			{journal} {\bibinfo  {journal} {Progress of Theoretical Physics Supplement}\
			}\textbf {\bibinfo {volume} {130}},\ \bibinfo {pages} {17} (\bibinfo {year}
			{1998})}\BibitemShut {NoStop}%
		\bibitem [{\citenamefont {Lee}\ \emph {et~al.}(2023)\citenamefont {Lee},
			\citenamefont {Kim}, \citenamefont {Park}, \citenamefont {Kim}, \citenamefont
			{Park},\ and\ \citenamefont {Lee}}]{Lee2023}%
		\BibitemOpen
		\bibfield  {author} {\bibinfo {author} {\bibfnamefont {S.}~\bibnamefont
				{Lee}}, \bibinfo {author} {\bibfnamefont {D.-K.}\ \bibnamefont {Kim}},
			\bibinfo {author} {\bibfnamefont {J.-M.}\ \bibnamefont {Park}}, \bibinfo
			{author} {\bibfnamefont {W.~K.}\ \bibnamefont {Kim}}, \bibinfo {author}
			{\bibfnamefont {H.}~\bibnamefont {Park}},\ and\ \bibinfo {author}
			{\bibfnamefont {J.~S.}\ \bibnamefont {Lee}},\ }\bibfield  {title} {\bibinfo
			{title} {Multidimensional entropic bound: Estimator of entropy production for
				langevin dynamics with an arbitrary time-dependent protocol},\ }\href
		{https://doi.org/https://doi.org/10.1103/PhysRevResearch.5.013194} {\bibfield
			{journal} {\bibinfo  {journal} {Physical Review Research}\ }\textbf
			{\bibinfo {volume} {5}},\ \bibinfo {pages} {013194} (\bibinfo {year}
			{2023})}\BibitemShut {NoStop}%
		\bibitem [{\citenamefont {Dechant}\ and\ \citenamefont
			{Sasa}(2018)}]{Dechant2018}%
		\BibitemOpen
		\bibfield  {author} {\bibinfo {author} {\bibfnamefont {A.}~\bibnamefont
				{Dechant}}\ and\ \bibinfo {author} {\bibfnamefont {S.-i.}\ \bibnamefont
				{Sasa}},\ }\bibfield  {title} {\bibinfo {title} {Entropic bounds on currents
				in langevin systems},\ }\href
		{https://doi.org/https://doi.org/10.1103/PhysRevE.97.062101} {\bibfield
			{journal} {\bibinfo  {journal} {Physical Review E}\ }\textbf {\bibinfo
				{volume} {97}},\ \bibinfo {pages} {062101} (\bibinfo {year}
			{2018})}\BibitemShut {NoStop}%
		\bibitem [{\citenamefont {Spinney}\ and\ \citenamefont
			{Ford}(2012)}]{Spinney2012}%
		\BibitemOpen
		\bibfield  {author} {\bibinfo {author} {\bibfnamefont {R.~E.}\ \bibnamefont
				{Spinney}}\ and\ \bibinfo {author} {\bibfnamefont {I.~J.}\ \bibnamefont
				{Ford}},\ }\bibfield  {title} {\bibinfo {title} {Entropy production in full
				phase space for continuous stochastic dynamics},\ }\href
		{https://doi.org/10.1103/PhysRevE.85.051113} {\bibfield  {journal} {\bibinfo
				{journal} {Phys. Rev. E}\ }\textbf {\bibinfo {volume} {85}},\ \bibinfo
			{pages} {051113} (\bibinfo {year} {2012})}\BibitemShut {NoStop}%
		\bibitem [{\citenamefont {Kwon}\ \emph {et~al.}(2024)\citenamefont {Kwon},
			\citenamefont {Park}, \citenamefont {Lee},\ and\ \citenamefont
			{Baek}}]{Kwon2024}%
		\BibitemOpen
		\bibfield  {author} {\bibinfo {author} {\bibfnamefont {E.}~\bibnamefont
				{Kwon}}, \bibinfo {author} {\bibfnamefont {J.-M.}\ \bibnamefont {Park}},
			\bibinfo {author} {\bibfnamefont {J.~S.}\ \bibnamefont {Lee}},\ and\ \bibinfo
			{author} {\bibfnamefont {Y.}~\bibnamefont {Baek}},\ }\bibfield  {title}
		{\bibinfo {title} {Unified hierarchical relationship between thermodynamic
				tradeoff relations},\ }\href {https://doi.org/10.1103/PhysRevE.110.044131}
		{\bibfield  {journal} {\bibinfo  {journal} {Phys. Rev. E}\ }\textbf {\bibinfo
				{volume} {110}},\ \bibinfo {pages} {044131} (\bibinfo {year}
			{2024})}\BibitemShut {NoStop}%
		\bibitem [{sup()}]{supp_NC}%
		\BibitemOpen
		\href@noop {} {}\bibinfo {note} {See Supplementary Information.}\BibitemShut
		{Stop}%
		\bibitem [{\citenamefont {Brinkman}(1956)}]{Brinkman1956}%
		\BibitemOpen
		\bibfield  {author} {\bibinfo {author} {\bibfnamefont {H.}~\bibnamefont
				{Brinkman}},\ }\bibfield  {title} {\bibinfo {title} {Brownian motion in a
				field of force and the diffusion theory of chemical reactions},\ }\href
		{https://doi.org/https://doi.org/10.1016/S0031-8914(56)80006-2} {\bibfield
			{journal} {\bibinfo  {journal} {Physica}\ }\textbf {\bibinfo {volume} {22}},\
			\bibinfo {pages} {29} (\bibinfo {year} {1956})}\BibitemShut {NoStop}%
		\bibitem [{\citenamefont {Seifert}(2005)}]{Seifert2005}%
		\BibitemOpen
		\bibfield  {author} {\bibinfo {author} {\bibfnamefont {U.}~\bibnamefont
				{Seifert}},\ }\bibfield  {title} {\bibinfo {title} {Entropy production along
				a stochastic trajecttory and an integral fluctuation theorem},\ }\href
		{https://doi.org/10.1103/PhysRevLett.95.040602} {\bibfield  {journal}
			{\bibinfo  {journal} {Phys. Rev. Lett.}\ }\textbf {\bibinfo {volume} {95}},\
			\bibinfo {pages} {040602} (\bibinfo {year} {2005})}\BibitemShut {NoStop}%
		\bibitem [{\citenamefont {Seifert}(2012)}]{Seifert2012}%
		\BibitemOpen
		\bibfield  {author} {\bibinfo {author} {\bibfnamefont {U.}~\bibnamefont
				{Seifert}},\ }\bibfield  {title} {\bibinfo {title} {Stochastic
				thermodynamics, fluctuation theorems and molecular machines},\ }\href
		{https://doi.org/10.1088/0034-4885/75/12/126001} {\bibfield  {journal}
			{\bibinfo  {journal} {Reports on Progress in Physics}\ }\textbf {\bibinfo
				{volume} {75}},\ \bibinfo {pages} {126001} (\bibinfo {year}
			{2012})}\BibitemShut {NoStop}%
		\bibitem [{\citenamefont {Rold{\'a}n}\ \emph {et~al.}(2014)\citenamefont
			{Rold{\'a}n}, \citenamefont {Mart{\'\i}nez}, \citenamefont {Dinis},\ and\
			\citenamefont {Rica}}]{Roldan2014}%
		\BibitemOpen
		\bibfield  {author} {\bibinfo {author} {\bibfnamefont {{\'E}.}~\bibnamefont
				{Rold{\'a}n}}, \bibinfo {author} {\bibfnamefont {I.~A.}\ \bibnamefont
				{Mart{\'\i}nez}}, \bibinfo {author} {\bibfnamefont {L.}~\bibnamefont
				{Dinis}},\ and\ \bibinfo {author} {\bibfnamefont {R.~A.}\ \bibnamefont
				{Rica}},\ }\bibfield  {title} {\bibinfo {title} {Measuring kinetic energy
				changes in the mesoscale with low acquisition rates},\ }\href
		{https://doi.org/10.1063/1.4882419} {\bibfield  {journal} {\bibinfo
				{journal} {Applied physics letters}\ }\textbf {\bibinfo {volume} {104}}
			(\bibinfo {year} {2014})}\BibitemShut {NoStop}%
		\bibitem [{\citenamefont {Li}\ \emph {et~al.}(2010)\citenamefont {Li},
			\citenamefont {Kheifets}, \citenamefont {Medellin},\ and\ \citenamefont
			{Raizen}}]{Li2010}%
		\BibitemOpen
		\bibfield  {author} {\bibinfo {author} {\bibfnamefont {T.}~\bibnamefont
				{Li}}, \bibinfo {author} {\bibfnamefont {S.}~\bibnamefont {Kheifets}},
			\bibinfo {author} {\bibfnamefont {D.}~\bibnamefont {Medellin}},\ and\
			\bibinfo {author} {\bibfnamefont {M.~G.}\ \bibnamefont {Raizen}},\ }\bibfield
		{title} {\bibinfo {title} {Measurement of the instantaneous velocity of a
				brownian particle},\ }\href {https://doi.org/10.1126/science.1189403}
		{\bibfield  {journal} {\bibinfo  {journal} {Science}\ }\textbf {\bibinfo
				{volume} {328}},\ \bibinfo {pages} {1673} (\bibinfo {year}
			{2010})}\BibitemShut {NoStop}%
		\bibitem [{\citenamefont {Huang}\ \emph {et~al.}(2011)\citenamefont {Huang},
			\citenamefont {Chavez}, \citenamefont {Taute}, \citenamefont {Luki{\'c}},
			\citenamefont {Jeney}, \citenamefont {Raizen},\ and\ \citenamefont
			{Florin}}]{Huang2011}%
		\BibitemOpen
		\bibfield  {author} {\bibinfo {author} {\bibfnamefont {R.}~\bibnamefont
				{Huang}}, \bibinfo {author} {\bibfnamefont {I.}~\bibnamefont {Chavez}},
			\bibinfo {author} {\bibfnamefont {K.~M.}\ \bibnamefont {Taute}}, \bibinfo
			{author} {\bibfnamefont {B.}~\bibnamefont {Luki{\'c}}}, \bibinfo {author}
			{\bibfnamefont {S.}~\bibnamefont {Jeney}}, \bibinfo {author} {\bibfnamefont
				{M.~G.}\ \bibnamefont {Raizen}},\ and\ \bibinfo {author} {\bibfnamefont
				{E.-L.}\ \bibnamefont {Florin}},\ }\bibfield  {title} {\bibinfo {title}
			{Direct observation of the full transition from ballistic to diffusive
				brownian motion in a liquid},\ }\href {https://doi.org/10.1038/nphys1953}
		{\bibfield  {journal} {\bibinfo  {journal} {Nature Physics}\ }\textbf
			{\bibinfo {volume} {7}},\ \bibinfo {pages} {576} (\bibinfo {year}
			{2011})}\BibitemShut {NoStop}%
		\bibitem [{\citenamefont {Kheifets}\ \emph {et~al.}(2014)\citenamefont
			{Kheifets}, \citenamefont {Simha}, \citenamefont {Melin}, \citenamefont
			{Li},\ and\ \citenamefont {Raizen}}]{Kheifets2014}%
		\BibitemOpen
		\bibfield  {author} {\bibinfo {author} {\bibfnamefont {S.}~\bibnamefont
				{Kheifets}}, \bibinfo {author} {\bibfnamefont {A.}~\bibnamefont {Simha}},
			\bibinfo {author} {\bibfnamefont {K.}~\bibnamefont {Melin}}, \bibinfo
			{author} {\bibfnamefont {T.}~\bibnamefont {Li}},\ and\ \bibinfo {author}
			{\bibfnamefont {M.~G.}\ \bibnamefont {Raizen}},\ }\bibfield  {title}
		{\bibinfo {title} {Observation of brownian motion in liquids at short times:
				instantaneous velocity and memory loss},\ }\href
		{https://doi.org/10.1126/science.1248091} {\bibfield  {journal} {\bibinfo
				{journal} {Science}\ }\textbf {\bibinfo {volume} {343}},\ \bibinfo {pages}
			{1493} (\bibinfo {year} {2014})}\BibitemShut {NoStop}%
	\end{thebibliography}
	%
	
	%
	%
	\section*{Acknowledgment}
	The authors acknowledge the Korea Institute for Advanced Study for providing computing resources (KIAS Center for Advanced Computation Linux Cluster System). This research was supported by NRF Grants No.~2017R1D1A1B06035497 (H.P.), and individual KIAS Grants No.~PG064902 (J.S.L.), PG096601 (S. A.), and QP013601 (H.P.) at the Korea Institute for Advanced Study.
	
	%
	%
	\section*{Author contributions}
    S.A. and J.S.L. conceived the initial idea and derived the main results. S.A. performed the numerical simulations. S.A., J.S.L., and H.P. contributed to discussions and writing the manuscript.
	%
	%
	\section*{Competing interests}
	The author declares no competing interests.

\end{document}


\title{SUPPLEMENTARY INFORMATION FOR\texorpdfstring{\\}{\newline}Thermodynamic anomalies in overdamped systems with time-dependent temperature}
	\author{Shakul Awasthi}
	\email{shakul23@kias.re.kr}
	\affiliation{School of Physics, Korea Institute for Advanced Study, Seoul 02455, Korea}
	\author{Hyunggyu Park}
	\email{hgpark@kias.re.kr}
	\affiliation{Quantum Universe Center, Korea Institute for Advanced Study, Seoul 02455, Korea}
	\author{Jae Sung Lee}
	\email{jslee@kias.re.kr}
	\affiliation{School of Physics, Korea Institute for Advanced Study, Seoul 02455, Korea}
    \affiliation{Quantum Universe Center, Korea Institute for Advanced Study, Seoul 02455, Korea}
	
	\date{\today}
	
    \maketitle
	
\section{Brinkman's Hierarchy for Langevin system with time-dependent temperature}
\subsection{Derivation of the coupled equations for Brinkman's coefficients}    

The Fokker-Planck operator $\mathcal{L}_{\rm ud}$ of Eq.~(3) in the main text can be divided into two parts, reversible $\mathcal{L}^\text{rev}$ and irreversible $\mathcal{L}^\text{irr}$ ones defined as
	\begin{align}
		\mathcal{L}^\text{rev} \equiv -\frac{\partial}{\partial x}v - \frac{1}{m}\frac{\partial}{\partial v} f(x,\lambda_t)~,\quad \mathcal{L}^\text{irr}\equiv - \frac{1}{m}\frac{\partial}{\partial v}\left( -\gamma v - \frac{\gamma T(t)}{m}  \frac{\partial}{\partial v}\right)~.
	\end{align}
The irreversible operator $\mathcal{L}^\text{irr}$ can be transformed into the Hermitianized form $\bar{\mathcal{L}}^\text{irr}$ as follows:
\begin{align} \label{eqS:irr_op_her}
    \bar{\mathcal{L}}^{\rm irr} \equiv e^{\Phi/2} \mathcal{L}^{\rm irr} e^{-\Phi/2},~~{\rm where }~ \Phi \equiv \frac{mv^2}{2T(t)}~.
\end{align}
By defining the creation $\hat{a}^\dagger$ and annihilation $\hat{a}$ operators as
\begin{align}
    \hat{a}^\dagger \equiv \frac{1}{2} \sqrt{\frac{m}{T(t)}}v - \sqrt{\frac{T(t)}{m}}\frac{\partial}{\partial v}~,~~\hat{a} \equiv \frac{1}{2} \sqrt{\frac{m}{T(t)}}v + \sqrt{\frac{T(t)}{m}}\frac{\partial}{\partial v}~,
\end{align}
$\bar{\mathcal{L}}^{\rm irr}$ can be written as
\begin{align} \label{eqS:irr_op_her_1}
    \bar{\mathcal{L}}^{\rm irr} = -\frac{\gamma}{m} \hat{a}^\dagger \hat{a}.
\end{align}
As Eq.~\eqref{eqS:irr_op_her_1} is identical to that of the harmonic oscillator, the $n$th eigenfunction of $\bar{\mathcal{L}}^{\rm irr}$ is given by
\begin{align} \label{eqS:psin-def}
    \psi_n  = \frac{1}{\sqrt{2^n n!} } \psi_0 H_n \left(\sqrt{\frac{m}{2T(t)}} v \right),
\end{align}
where $\psi_0 = N e^{-\Phi/2}$ with the normalization factor $N \equiv \left[ 2 \pi T(t)/m\right]^{-1/4}$ and $H_n(x)$ denotes the Hermite polynomial, i.e., $H_0(x)=1,~ H_1(x)=2x,~ H_2(x)=4x^2-2, \cdots$. Thus, $\bar{\mathcal{L}}^{\rm irr} \psi_n = -\varepsilon_n \psi_n$ with the eigenvalue $\varepsilon_n = \gamma n/m$. The reversible operator $\mathcal{L}^{\rm rev}$ can be similarly transformed as
\begin{align} \label{eqS:rev_op_her}
    \bar{\mathcal{L}}^{\rm rev} \equiv e^{\Phi/2} \mathcal{L}^{\rm rev} e^{-\Phi/2} = -\hat{b}\hat{a} - \hat{b}^\prime \hat{a}^\dagger,
\end{align}
where $\hat{b} \equiv \sqrt{T(t)/m} \partial_x$ and $\hat{b}^\prime \equiv \sqrt{T(t)/m}\partial_x - \sqrt{m/T(t)} f(x,\lambda_t)/m$.

Up to this point, the derivation procedure is exactly the same as that of the conventional Brinkman's hierarchy method~\cite{Brinkman1956,Risken1996}. However, in the remaining steps, the time dependence of $\psi_n$ introduces additional mathematical complexity, as presented below.  
By multiplying $N^{-1}e^{\Phi/2}$ to both sides of Eq.~(2) in the main text and defining $\bar{P}_{\rm ud} \equiv \psi_0^{-1}  P_{\rm ud} (x,v,t)$, we obtain the following:
\begin{align} \label{eqS:FP_hermitianized}
    \partial_t \bar{P}_{\rm ud} - (\partial_t \psi_0^{-1}) P_{\rm ud} = (\bar{\mathcal{L}}^{\rm rev} + \bar{\mathcal{L}}^{\rm irr}) \bar{P}_{\rm ud}. 
\end{align}
We note that the term $(\partial_t \psi_0^{-1}) P_{\rm ud}$ in Eq.~\eqref{eqS:FP_hermitianized} is absent in the conventional Brinkman's hierarchy calculation~\cite{Brinkman1956,Risken1996}, where the temperature has no dependence on time. Explicit calculation leads to 
\begin{equation} \label{eqS:J}
    (\partial_t \psi_0^{-1}) P_{\rm ud} = \mathcal{J} \bar{P}_{\rm ud},~~{\rm where } \mathcal{J} \equiv \frac{\dot{T}(t)}{4T(t)^2} (T - mv^2).
\end{equation}

Now, we expand $\bar{P}_{\rm ud}$ in terms of the eigenfunction $\psi_n$ as
\begin{equation} \label{Pbar-exp}
    \bar{P}_{\rm ud} = \sum_{n=0}^{\infty} c_n(x,t) \psi_n.
\end{equation}
Note that the dependence of $\bar{P}_\text{ud}$ on the position and velocity is split into purely velocity-dependent functions $\psi_n$ and purely position-dependent coefficients $c_n$. In addition, the $\gamma$-dependence of the distribution is contained solely in the coefficients $c_n$. $\psi_n$ depends on time due to the time-varying temperature $T(t)$. Plugging Eqs.~\eqref{eqS:rev_op_her}, ~\eqref{eqS:J}, and \eqref{Pbar-exp} into Eq.~\eqref{eqS:FP_hermitianized} leads to
\begin{align}\label{EqS:cn-psin-eq-1}
    \sum_{n=0}^\infty (\partial_t c_n) \psi_n + \sum_{n=0}^\infty c_n \partial_t \psi_n = & - \sum_{n=0}^\infty \sqrt{\frac{(n+1)T(t)}{m}} \partial_x c_{n+1} \psi_n + \frac{\sqrt{n}}{\sqrt{T(t)m}} \sum_{n=1}^\infty \left[ f(x,\lambda_t) -T(t) \partial_x   \right] c_{n-1} \psi_n \nonumber \\
    &-\frac{\gamma}{m} \sum_{n=0}^\infty c_n n \psi_n + \sum_{n=0}^\infty \mathcal{J} c_n \psi_n
\end{align}
In Eq.~\eqref{EqS:cn-psin-eq-1}, $\partial_t \psi_n$ can be calculated as 
\begin{align} \label{EqS:dt-psin}
    \partial_t \psi_n &= \frac{1}{\sqrt{2^n n!}} \left[ (\partial_t \psi_0 ) H_n \left( \sqrt{\frac{m}{2T}v} \right) +  \psi_0  \partial_t H_n \left( \sqrt{\frac{m}{2T}v} \right) \right] \nonumber \\
    & = \frac{1}{\sqrt{2^n n!}} \left[ -\mathcal{J} H_n \left( \sqrt{\frac{m}{2T}v} \right) -nv \sqrt{\frac{m}{2T}} \left(\frac{\dot{T}}{T} \right) \psi_0  H_{n-1} \left( \sqrt{\frac{m}{2T}v} \right) \right] \nonumber \\
    &=- \mathcal{J} \psi_n - \sqrt{\frac{nm}{4T}}\frac{\dot{T}}{T}v\psi_{n-1}~.
\end{align}
Substituting Eq.~\eqref{EqS:dt-psin} into Eq.~\eqref{EqS:cn-psin-eq-1}, we obtain
\begin{align}\label{EqS:cn-psin-eq-2}
    \sum_{n=0}^\infty (\partial_t c_n) \psi_n   
    = & - \sum_{n=0}^\infty \sqrt{\frac{(n+1)T(t)}{m}} \partial_x c_{n+1} \psi_n + \frac{\sqrt{n}}{\sqrt{T(t)m}} \sum_{n=1}^\infty \left[ f(x,\lambda_t) -T(t) \partial_x   \right] c_{n-1} \psi_n -\frac{\gamma}{m} \sum_{n=0}^\infty c_n n \psi_n \nonumber \\
    &+ 2\sum_{n=0}^\infty \mathcal{J} c_n \psi_n +  \sum_{n=0}^\infty \sqrt{\frac{(n+1)m}{4T}}\frac{\dot{T}}{T}vc_{n+1}\psi_n~.
\end{align}
To extract the hierarchical relationship between $c_n$s from Eq.~\eqref{EqS:cn-psin-eq-2}, the following identities are necessary:
\begin{equation}
\begin{aligned} \label{EqS:iden}
    &\int_{-\infty}^{\infty} dv~ \psi_n(v) \psi_l(v) = \delta_{n,l}~,\\ 
    &\int_{-\infty}^{\infty}dv \mathcal{J} \psi_n(v) \psi_l(v) = \frac{\dot{T}}{4 T} \left[ -2 l\delta_{n,l} - \sqrt{l(l-1)}\delta_{n,l-2} - \sqrt{(l+1)(l+2)}\delta_{n,l+2}\right]~,\\ 
    &\int_{-\infty}^{\infty}dv~ v \psi_n(v) \psi_l(v) = \sqrt{\frac{2 T}{m}}\left(\sqrt{\frac{l}{2}}\delta_{n,l-1} + \sqrt{\frac{l+1}{2}}\delta_{n,l+1}\right)~,
\end{aligned}
\end{equation}
where $\delta_{n,l}$ is the Kronecker delta function. To obtain the second and the third equalities in Eq.~\eqref{EqS:iden}, the following relation is used
\begin{align} \label{eqS:v_psi_n}
    v \psi_n(v) = \sqrt{\frac{2T}{m}} \left[ \sqrt{\frac{n+1}{2}} \psi_{n+1}(v) + \sqrt{\frac{n}{2}} \psi_{n-1}(v)  \right],
\end{align}
which can be derived using the recurrence relation of the Hermite polynomial, i.e., $xH_n(x) = \frac{1}{2} H_{n+1}(x) + n H_{n-1}(x)$.

Multiplying Eq.~\eqref{EqS:cn-psin-eq-2} by $\psi_l$ and integrating over the velocity variable yields the hierarchical relationship of the coefficients. For instance, multiplying Eq.~\eqref{EqS:cn-psin-eq-2} by $\psi_0$, integrating over the velocity variable, and then using Eqs.~\eqref{EqS:iden}, we obtain
\begin{equation}\label{c0-eqn}
    \partial_t c_0= -\sqrt{\frac{T(t)}{m}}\partial_x c_1~.
\end{equation}
Similarly, by multiplying Eq.~\eqref{EqS:cn-psin-eq-2} by $\psi_1$, we have
\begin{equation} \label{c1-eqn}
    \partial_t c_1 = - \sqrt{\frac{2T(t)}{m}}\partial_x c_2 + \frac{1}{\sqrt{m T(t)}} [ f(x,\lambda_t) - T(t)\partial_x ]c_0 - \frac{\gamma}{m}c_1 - \frac{\dot{T}(t)}{2 T(t)}c_1  ~.
\end{equation}
For general $\psi_l$ with $l \geq 2$, the resulting equation, after substituting $l$ with $n$, is 
\begin{align}
\label{cn-eqn}
    \partial_t c_n = -\sqrt{\frac{(n+1)T(t)}{m}}\partial_x c_{n+1} + \frac{\sqrt{n}}{\sqrt{m T(t)}} \left[ f(x,\lambda_t) -T(t)\partial_x \right] c_{n-1} - \frac{n \gamma}{m}c_n - \frac{\dot{T}(t)}{2 T(t)} \left( \sqrt{n(n-1)} c_{n-2} +n c_n  \right)~.
\end{align}
As seen in Eq.~\eqref{cn-eqn}, the evolution of the $n$th coefficient is coupled to the $(n+1)$, $(n-1)$ and $(n-2)$th coefficients. Therefore, to solve the coupled equations, all coefficients must be determined simultaneously. However, in the large $\gamma$ or small $m$ limit, we can deduce that the contribution of the higher-order coefficients becomes negligibly small. Thus, only the first few coefficients are needed in the limit.

Using the orthonormality of the eigenfunctions $\psi_n$, i.e., $\int_{-\infty}^\infty dv~ \psi_n \psi_m = \delta_{nm}$, we can show that $c_0$ is identical to the marginal distribution of $P_{\rm ud}$ as follows: 
\begin{equation} \label{eqS:c0_prob}
     P_{\rm od}(x,t) = \int_{-\infty}^{\infty} dv P_{\rm ud}(x,v,t)      = \int_{-\infty}^{\infty} dv~ \psi_0 \bar{P}(x,v,t)      = \sum_{n=0}^\infty c_n \int_{-\infty}^{\infty} dv~ \psi_0 \psi_n = c_0 ~.
\end{equation}

\subsection{High-viscosity regime: \texorpdfstring{$z=0$}{z=0}}  
Let us first consider the large-$\gamma$ regime, which is one of the typical conditions leading to overdamped dynamics. The other conditions, the small-$m$ and intermediate regimes, will be discussed in Secs.~\ref{secS:small_mass} and \ref{secS:intermediate_regime}. 
As the probability distribution cannot diverge in the large-$\gamma$ limit, $c_n$ can be expressed as inverse powers of $\gamma$, as follows:
\begin{equation}\label{cn-exp-all}
    c_n(x,t) = \sum_{j=0}^{\infty} c_n^{(j)}(x,t)\gamma^{-j}~.
\end{equation}
Here, $c_0$ is considered to be of order $O(\gamma^{0})$ due to the normalization condition $\int_{-\infty}^\infty dx ~ c_0 =1$ (see Eq.~\eqref{eqS:c0_prob}), implying $c_0^{(0)} \neq 0$. Substituting Eq.~\eqref{cn-exp-all} into Eq.~\eqref{c1-eqn} and collecting the terms of the same order in $\gamma$, we find that $c_1^{(0)} = 0$. Therefore, $c_1$ is of order $O(\gamma^{-1})$. For $n \geq 2$, from Eq.~\eqref{cn-eqn}, it is clear that $O(c_n) = O(c_{n-2}/\gamma)$. Therefore, we generally deduce that all $c_n$s are of order $O(\gamma^{\lfloor{-n/2}\rfloor})$, where $\lfloor \cdots \rfloor$ represents the floor function.

The calculation of heat in the overdamped approximation requires the terms up to the order of $\gamma^{-2}$, as explained in Sec~\ref{secS:heat}. Therefore, we aim to calculate all quantities up to that order. Then, the relevant $c_n$ coefficients $(n=0,1,\cdots,4)$ can be expanded as
\begin{align}\label{cn-exp}
    \begin{split}
        &c_0 = c_0^{(0)} +\frac{1}{\gamma}c_0^{(1)} + \frac{1}{\gamma^2}c_0^{(2)}, \quad \quad 
        c_1 =  \frac{1}{\gamma}c_1^{(1)} + \frac{1}{\gamma^2}c_1^{(2)},\\
        &c_2 = \frac{1}{\gamma}c_2^{(1)} + \frac{1}{\gamma^2}c_2^{(2)},\quad\quad 
        c_3 = \frac{1}{\gamma^2}c_3^{(2)},\quad\quad 
        c_4 = \frac{1}{\gamma^2}c_4^{(2)}~.
    \end{split}
\end{align} 
All $c_n$ with $n\geq5$ are of higher order than $O(\gamma^{-2})$. Inserting Eq.~\eqref{cn-exp} into the coupled equations~\eqref{c0-eqn}, \eqref{c1-eqn}, and \eqref{cn-eqn}, we can construct the hierarchical relationship between $c_n^{(i)}$. 
First, plugging the expanded expressions of $c_0$ and $c_1$ into Eq.~\eqref{c0-eqn} and arranging the terms in the same order, we obtain
\begin{align}
    \partial_t c_0^{(0)} = 0,\quad \partial_t c_0^{(1)} =  -\sqrt{\frac{T(t)}{m}}\partial_x c_1^{(1)},\quad \partial_t c_0^{(2)} =  -\sqrt{\frac{T(t)}{m}}\partial_x c_1^{(2)}~.
\end{align}
Next, collecting the same order terms in Eq.~\eqref{c1-eqn}, we have the following two equations:
\begin{subequations}
\begin{align}
    0 &= \frac{1}{\sqrt{m T(t)}} [ f(x,\lambda_t)  - T(t) \partial_x ] c_0^{(0)}  - \frac{1}{m} c_1^{(1)}  \label{eqS:n1a}\\
    \partial_t c_1^{(1)} &=- \sqrt{\frac{2T(t)}{m}}\partial_x c_2^{(1)} +  \frac{1}{\sqrt{m T(t)}} [ f(x,\lambda_t)  - T(t) \partial_x ] c_0^{(1)} - \frac{1}{m} c_1^{(2)} -\frac{\dot{T}(t)}{2T(t)} c_1^{(1)}   \label{eqS:n1b}
\end{align}
\end{subequations}
From Eq.~\eqref{eqS:n1a}, $c_1^{(1)}$ becomes
\begin{align} \label{eqS:c_1_1}
    c_1^{(1)} = \sqrt{\frac{m}{T(t)}} \left[f(x,\lambda_t) - T(t) \partial_x \right] c_0^{(0)}~.
\end{align}
Substituting Eq.\eqref{eqS:c_1_1} and the expression of $c_2^{(1)}$ from Eq.~\eqref{c2n-eqns_a} into Eq.~\eqref{eqS:n1b}, we obtain $c_1^{(2)}$ as
\begin{align}
    c_1^{(2)} = \sqrt{\frac{m}{T(t)}} \left[f(x,\lambda_t) - T(t) \partial_x \right] c_0^{(1)}+ \frac{3 m}{2}\sqrt{\frac{m}{T(t)}} \dot{T}(t)\partial_x c_0^{(0)}  - \frac{m\sqrt{m}}{\sqrt{T(t)}} \frac{\partial f(x,\lambda_t)}{\partial \lambda_t}  \dot{\lambda}_t c_0^{(0)}  ~.
\end{align}
From Eq.~\eqref{cn-eqn}, the equation for $n=2$ is written as
\begin{equation}
    \partial_t c_2 = - \sqrt{\frac{3T(t)}{m}}\partial_x c_3 + {\frac{\sqrt{2}}{\sqrt{mT(t)}}}\left[f(x,\lambda_t) - T\partial_x \right]c_1 -\frac{2\gamma}{m} c_2 - \frac{\dot{T}(t)}{2 T(t)}\left(\sqrt{2}c_0 + 2 c_2\right) ~.
\end{equation}
Then, following the similar procedure used for obtaining $c_1^{(1)}$ and $c_1^{(2)}$, we get 
\begin{subequations}
\begin{align}
    c_2^{(1)} &= -\frac{m}{2 \sqrt{2}}\left(\frac{\dot{T}}{T}\right)c_0^{(0)}, \label{c2n-eqns_a} \\ 
    c_2^{(2)} &= \frac{m f^2}{\sqrt{2}T}c_0^{(0)}  -\sqrt{2} m f \partial_xc_0^{(0)} - \frac{m}{\sqrt{2}} \frac{\partial f}{\partial x}c_0^{(0)} + \frac{m}{\sqrt{2}}T \partial_x^2c_0^{(0)}   - \frac{m}{2 \sqrt{2}}\left(\frac{\dot{T}}{T}\right)c_0^{(1)} + \frac{m^2}{4 \sqrt{2}}\left(\frac{\ddot{T}}{T}\right)c_0^{(0)}~. \label{c2n-eqns_b}
\end{align} 
\end{subequations}
Similarly for $n=3$, we get the following equation and $c_3^{(2)}$:
\begin{align}
    &\partial_t c_3 = - 2\sqrt{\frac{T(t)}{m}}\partial_x c_4  + \frac{\sqrt{3}}{\sqrt{mT(t)}} \left[ f(x,\lambda_t) - T(t)\partial_x \right] c_2   - \frac{3 \gamma}{m}c_3 - \frac{\dot{T}(t)}{2 T(t)}\left(\sqrt{6} c_1 + 3 c_3 \right)~. \\
    &c_3^{(2)} = -\frac{\sqrt{6}}{4} \left(\frac{m}{T(t)} \right)^{\frac{3}{2}} \dot{T}(t) [f(x,\lambda_t)-T(t) \partial_x] c_0^{(0)}~.
\end{align}
Finally, equation for $n=4$ and $c_4^{(2)}$ are given by 
\begin{align}
    &\partial_t c_4 = - \sqrt{\frac{5 T(t)}{m}}\partial_x c_5 + \frac{2}{\sqrt{m T(t)}} [f(x,\lambda_t) - T(t) \partial_x ] c_3 - \frac{4 \gamma}{m}c_4 - \frac{\dot{T}(t)}{2 T(t)}\left(2\sqrt{3} c_2 + 4 c_4 \right)  ~,\\
    &c_4^{(2)} = \frac{1}{8}\sqrt{ \frac{3}{2} }m^2\left(\frac{\dot{T}}{T}\right)^2c_0^{(0)}~.
\end{align}
We have obtained all the coefficients up to the order of $\gamma^{-2}$ in terms of $c_0$ and its spatial derivatives.  

\subsection{Small-mass regime: \texorpdfstring{$z=1/2$}{z=1/2}}  \label{secS:small_mass}
Next, we consider the small-mass limit, which is another condition leading to overdamped dynamics. In this limit, the coefficient $c_n$ can be expanded in powers of the mass $m$. Carefully examining the governing equation~\eqref{cn-eqn}, we can deduce that $O(c_{n+1}) = O(\sqrt{m} c_n)$. Therefore, $c_n$ can be expanded as  
\begin{equation}\label{cn-m-exp-all}
    c_n(x,t) = \sum_{j=0}^{\infty} c_n^{(j)}(x,t)(\sqrt{m})^{j}~. 
\end{equation} 
Since $c_0$ is of order $O(m^0)$, the order of $c_n$ is $O(\sqrt{m}^n)$. Applying a similar procedure as in the large-$\gamma$ regime, we can find the hierarchical relationship of $c_n^{(j)}$. The results up to order $O(m)$ are summarized as follows:
For $c_0$, the results are
\begin{align}\label{c0-m-exp}
    \partial_t c_0^{(0)} = -\sqrt{T(t)}\partial_x c_1^{(1)},~~ \partial_t c_0^{(1)} = -\sqrt{T(t)}\partial_x c_1^{(2)}. 
\end{align} 
$c_1^{(1)}$ and $c_1^{(2)}$ are given by
\begin{align} \label{c1-m-exp}  
    c_1^{(1)} = \frac{1}{\gamma \sqrt{T}}\left[f - T \partial_x \right] c_0^{(0)}~,\quad c_1^{(2)} = \frac{1}{\gamma \sqrt{T}}\left[f - T \partial_x \right] c_0^{(1)}~.
\end{align}
Finally, the expression of $c_2^{(2)}$ is 
\begin{align}\label{c22-m-exp}
     c_2^{(2)} = \frac{1}{\sqrt{2}\gamma^2 T}\left[ f^2 - T(\partial_x f) - 2fT \partial_x + T^2 \partial_x^2 \right] c_0^{(0)} -\frac{\sqrt{2}\dot{T}}{4\gamma T} c_0^{(0)} ~.
\end{align}

\subsection{Dimensionless coupled equations for Brinkman's coefficients with two-scaling parameters} \label{secS:Brinkman_two_scaling}

As shown in the previous subsections, high-viscosity and the small-mass limits result in different expressions for the Brinkman's coefficients. To address this in a systematic way, here, we present a unified perturbative scheme, Brinkman's hierarchy with two-scaling parameters, capable of exploring not only these two limits but also the intermediate regimes. Note that the same derivation is also presented in Appendix of the main text.

To achieve this, we convert Eq.~\eqref{cn-eqn} into a dimensionless form by introducing the characteristic time and length scales of the overdamped system, denoted as $\tau_\text{od}$ and $l_\text{od}$, respectively. 
Using these, we define the dimensionless time, position, and $n$th coefficient as $\bar{t} \equiv t/\tau_\text{od}$, $\bar{x} \equiv x/l_\text{od}$, and $\bar{c}_n \equiv l_\text{od} c_n$, respectively. Additionally, we introduce a dimensionless temperature $\bar{T}(t) \equiv T(t)/T_0$, where $T_0$ represents the typical energy scale of the system.
These definitions allow us to specify the typical velocities of the underdamped system, $V_\text{ud} \equiv \sqrt{T_0/m}$, and the overdamped system, $V_\text{od} \equiv l_\text{od}/\tau_\text{od}$. Using the quantities defined thus far, we can rewrite Eq.~\eqref{cn-eqn} in a dimensionless form as follows:
\begin{align}
\label{eqA:cbn-eqn_dimless}
    \partial_{\bar{t}} \bar{c}_n = -\nu \sqrt{(n+1)\bar{T}(t)}\partial_{\bar{x}} \bar{c}_{n+1} + \nu \sqrt{n\bar{T}(t)}\left[ \frac{\bar{f}(x,\lambda_t)}{\bar{T}(t)} -\partial_{\bar{x}} \right] \bar{c}_{n-1} - \tau n \bar{c}_n - \frac{\dot{\bar{T}}(t)}{2 \bar{T}(t)} \left( \sqrt{n(n-1)} \bar{c}_{n-2} +n \bar{c}_n  \right)~,
\end{align}
where $\nu \equiv V_\text{ud}/V_\text{od}$, $\tau \equiv \tau_\text{od}/\tau_{\rm r}$, $\dot{\bar{T}}=d\bar{T}/d\bar{t}$, and $\bar{f} \equiv (l_\text{od}/T_0) f$ is the dimensionless force.

\subsection{Intermediate regime: \texorpdfstring{$0<z<1/2$}{0<z<1/2}}
\label{secS:intermediate_regime}

Substituting $\nu = \tau^z$ into Eq.~\eqref{eqA:cbn-eqn_dimless} for $n=0,1$, and $2$ leads to
\begin{align}
    &\partial_{\bar{t}} \bar{c}_0 = -\tau^z \sqrt{\bar{T}(t)}\partial_{\bar{x}} \bar{c}_{1} \;, \label{cbn-eqn_dimless_0} \\
    &\partial_{\bar{t}} \bar{c}_1 = -\tau^z \sqrt{2\bar{T}(t)}\partial_{\bar{x}} \bar{c}_{2} + \tau^z \sqrt{\bar{T}(t)}\left[ \frac{\bar{f}(x,\lambda_t)}{\bar{T}(t)} -\partial_{\bar{x}} \right] \bar{c}_{0} - \tau  \bar{c}_1 - \frac{\dot{\bar{T}}(t)}{2 \bar{T}(t)} \bar{c}_1 ~,\label{cbn-eqn_dimless_1} \\
    &\partial_{\bar{t}} \bar{c}_2 = -\tau^z \sqrt{3\bar{T}(t)}\partial_{\bar{x}} \bar{c}_{3} + \tau^z \sqrt{2\bar{T}(t)}\left[ \frac{\bar{f}(x,\lambda_t)}{\bar{T}(t)} -\partial_{\bar{x}} \right] \bar{c}_{1} - 2\tau  \bar{c}_2 - \frac{\dot{\bar{T}}(t)}{2 \bar{T}(t)} \left( \sqrt{2} \bar{c}_{0} +2 \bar{c}_2  \right)~. \label{cbn-eqn_dimless_2}
\end{align}
For $0 < z < 1/2$, collecting the leading-order terms for $n=1$ and $n=2$ result in
\begin{align}
    &\bar{c}_1 \approx \tau^{z-1} \sqrt{\bar{T}(t)}\left( \bar{f}/\bar{T} - \partial_{\bar{x}}\right)\bar{c}_0\;.\label{eqA:FPbar-eqn1} \\
    & \bar{c}_2 \approx  - \frac{\dot{\bar{T}}(t)}{2\sqrt{2} \bar{T}(t) \tau} \bar{c}_{0}~. \label{eqA:FPbar-eqn2}
\end{align}
Therefore, the orders of the first three coefficients are $O(\bar c_0) = \tau^0$, $O(\bar c_1) = \tau^{-1+z}$, and $O(\bar c_2) = \tau^{-1}$. The relation for all higher-order coefficients with $n\geq 3$ is given by $O(\bar c_n) = O(\bar c_{n-2})/\tau$.  

As the value of $z$ is not fixed, calculating the higher-order terms is not straightforward. Therefore, we first write the expansion of $\bar c_n$ with the leading three terms as
\begin{align}
    \bar{c}_0 &= \bar{c}_0^{(0)} + \frac{1}{\tau^{\alpha_1}}  \bar{c}_0^{(1)} + \frac{1}{\tau^{\alpha_2}}  \bar{c}_0^{(2)}+ \cdots~,\label{cb0-exp}\\
    \bar{c}_1 &= \frac{\bar{c}_1^{(0)}}{\tau^{1-z}} + \frac{1}{\tau^{\beta_1}} \beta_1 \bar{c}_1^{(1)} + \frac{1}{\tau^{\beta_2}} \bar{c}_1^{(2)}+ \cdots~,\label{cb1-exp}\\
    \bar{c}_2 &= \frac{1}{\tau} \bar{c}_2^{(0)} + \frac{1}{\tau^{\gamma_1}} \bar{c}_2^{(1)} + \frac{1}{\tau^{\gamma_2}} \bar{c}_2^{(2)}+ \cdots~.\label{cb2-exp} \\
    \bar{c}_3 &= \frac{1}{\tau^{2-z}} \bar{c}_3^{(0)} + \frac{1}{\tau^{\omega_1}} \bar{c}_3^{(1)} + \frac{1}{\tau^{\omega_2}} \bar{c}_3^{(2)}+ \cdots~.\label{cb3-exp}
\end{align}
To determine the set of exponents $\{ \alpha_i, \beta_i, \gamma_i, \omega_i\}$, we substitute the above expansion into the corresponding governing equations. For instance, to obtain $ \alpha_i$, we substitute Eqs.~\eqref{cb0-exp} and \eqref{cb1-exp} into Eq.~\eqref{cbn-eqn_dimless_0}, yielding
\begin{equation} \label{eqS:alpha_determination}
	\partial_{\bar{t}} \left( \bar{c}_0^{(0)} + \frac{1}{\tau^{\alpha_1}} \bar{c}_0^{(1)} + \frac{1}{\tau^{\alpha_2}} \bar{c}_0^{(2)} \right) = 
	-\tau^z \sqrt{\bar{T}}  \partial_{\bar{x}} \left( \frac{\bar{c}_1^{(0)}}{\tau^{1-z}} + \frac{\bar{c}_1^{(1)}}{\tau^{\beta_1}} + \frac{\bar{c}_1^{(2)}}{\tau^{\beta_2}} \right)\;.
\end{equation}
Matching terms of Eq.~\eqref{eqS:alpha_determination} order by order, we obtain 
\begin{align}
    &\partial_{\bar{t}} \bar{c}_0^{(0)} = 0~, \\
	& \frac{1}{\tau^{\alpha_1}} \partial_{\bar{t}} \bar{c}_0^{(1)} = -\frac{1}{\tau^{1-2z}} \sqrt{\bar{T}} \partial_{\bar{x}} \bar{c}_1^{(0)}~, \label{cb01-eqn}\\
	&\frac{1}{\tau^{\alpha_2}} \partial_{\bar{t}} \bar{c}_0^{(2)} = -\frac{1}{\tau^{\beta_1-z}} \sqrt{\bar{T}} \partial_{\bar{x}}\bar{c}_1^{(1)}~.\label{cb02-eqn}
\end{align}
Thus, from Eq.~\eqref{cb01-eqn}, we find that $\alpha_1 = 1 - 2z$. Likewise, Eq.~\eqref{cb02-eqn} gives $\alpha_2 = \beta_1 - z$. More generally, we obtain $\alpha_{i+1} = \beta_i - z \text{ for } i \geq 1$. 

Now, substituting Eqs.~\eqref{cb0-exp}, \eqref{cb1-exp}, and \eqref{cb2-exp} into Eq.~\eqref{cbn-eqn_dimless_1}, we have
\begin{align} \label{eqS:c_1_matching}
    \partial_{\bar{t}} \left( \frac{\bar{c}_1^{(0)}}{\tau^{1-z}} + \frac{\bar{c}_1^{(1)}}{\tau^{\beta_1}} + \frac{\bar{c}_1^{(2)}}{\tau^{\beta_2}} \right) =&  -\tau^z \sqrt{2\bar{T}} \partial_{\bar{x}}\left( \frac{\bar{c}_2^{(0)}}{\tau} + \frac{\bar{c}_2^{(1)}}{\tau^{\gamma_1}} \right)+ \tau^z \sqrt{\bar{T}}\left(\frac{\bar{f}}{\bar{T}} - \partial_{\bar{x}}\right) \left( \bar{c}_0^{(0)} + \frac{\bar{c}_0^{(1)}}{\tau^{1-2z}} + \frac{\bar{c}_0^{(2)}}{\tau^{\alpha_2}} \right) \nonumber \\& - \tau \left( \frac{\bar{c}_1^{(0)}}{\tau^{1-z}} + \frac{\bar{c}_1^{(1)}}{\tau^{\beta_1}} + \frac{\bar{c}_1^{(2)}}{\tau^{\beta_2}} \right) - \frac{\dot{\bar{T}}}{2 \bar{T}} \left( \frac{\bar{c}_1^{(0)}}{\tau^{1-z}} + \frac{\bar{c}_1^{(1)}}{\tau^{\beta_1}} + \frac{\bar{c}_1^{(2)}}{\tau^{\beta_2}} \right)~.
\end{align}
Matching the leading order terms of Eq.~\eqref{eqS:c_1_matching} results in 
\begin{align} \label{eqA:FPbar-eqn1-0} 
    &\bar{c}_1^{(0)} = \sqrt{\bar{T}}\left( \frac{\bar{f}}{\bar{T}} - \partial_{\bar{x}}\right)\bar{c}_0^{(0)}\;.
\end{align}
Matching the next order terms of Eq.~\eqref{eqS:c_1_matching}, $\tau^z \sqrt{\bar{T}}\left(\bar{f}/\bar{T} - \partial_{\bar{x}}\right)   \bar{c}_0^{(1)}/\tau^{1-2z}$ and
$- \tau (  \bar{c}_1^{(1)} /\tau^{\beta_1}  )$, yields $\beta_1 = 2 - 3z$ and
\begin{equation}
    \bar{c}_1^{(1)} = \sqrt{\bar{T}}\left( \frac{\bar{f}}{\bar{T}} - \partial_{\bar{x}}\right)\bar{c}_0^{(1)}~. 
\end{equation}
Thus, $\alpha_2 = \beta_1 -z = 2 - 4z$. 

Finally, substituting Eqs.~\eqref{cb0-exp}, \eqref{cb1-exp}, \eqref{cb2-exp}, and \eqref{cb3-exp} into Eq.~\eqref{cbn-eqn_dimless_2} gives
\begin{align} \label{eqS:c2_expansion}
    \partial_{\bar{t}} \left( \frac{1}{\tau} \bar{c}_2^{(0)} + \frac{1}{\tau^{\gamma_1}} \bar{c}_2^{(1)} \right) =& - \tau^z \sqrt{3 \bar{T}} \partial_{\bar{x}} \left( \frac{1}{\tau^{2-z}} \bar{c}_3^{(0)} + \frac{1}{\tau^{\omega_1}} \bar{c}_3^{(1)} \right) + \tau^z \sqrt{2 \bar{T}} \left( \frac{\bar{f}}{\bar{T}}-\partial_{\bar{x}} \right) \left( \frac{\bar{c}_1^{(0)}}{\tau^{1-z}} + \frac{\bar{c}_1^{(1)}}{\tau^{2-3z}} + \frac{\bar{c}_1^{(2)}}{\tau^{\beta_2}} \right) \nonumber \\ &- 2 \tau \left( \frac{1}{\tau} \bar{c}_2^{(0)} + \frac{1}{\tau^{\gamma_1}} \bar{c}_2^{(1)} \right) - \frac{\dot{\bar{T}}}{2 \bar{T}} \sqrt{2} \left( \bar{c}_0^{(0)} + \frac{1}{\tau^{1-2z}} \bar{c}_0^{(1)} + \frac{1}{\tau^{2-4z}} \bar{c}_0^{(2)} \right) - \frac{\dot{\bar{T}}}{\bar{T}} \left( \frac{1}{\tau} \bar{c}_2^{(0)} + \frac{1}{\tau^{\gamma_1}} \bar{c}_2^{(1)} \right)~.
\end{align}
The leading-order terms of Eq.~\eqref{eqS:c2_expansion}, $- 2 \tau \left( \frac{1}{\tau} \bar{c}_2^{(0)}  \right)$ and $- \frac{\dot{\bar{T}}}{2 \bar{T}} \sqrt{2} \bar{c}_0^{(0)}$ ,  are matched as  
\begin{align} \label{eqS:bar_c2_0}
    & \bar{c}_2^{(0)} =  - \frac{\dot{\bar{T}}(t)}{2\sqrt{2} \bar{T}(t) } \bar{c}_{0}^{(0)}~. 
\end{align}
By matching the next order terms of Eq.~\eqref{eqS:c2_expansion}, $\tau^z \sqrt{2 \bar{T}} \left( \frac{\bar{f}}{\bar{T}}-\partial_{\bar{x}} \right) \left( \frac{\bar{c}_1^{(0)}}{\tau^{1-z}} \right)$, $- 2 \tau \left( \frac{1}{\tau^{\gamma_1}} \bar{c}_2^{(1)} \right)$, and $- \frac{\dot{\bar{T}}}{2 \bar{T}} \sqrt{2} \left( \frac{1}{\tau^{1-2z}} \bar{c}_0^{(1)} \right)$, 
we find that $\gamma_1 = 2 - 2z$ and 
\begin{equation}\label{c21-feqn}
    \bar{c}_2^{(1)} = \frac{\bar{T}}{\sqrt{2}}\left(\frac{\bar{f}}{\bar{T}} - \partial_{\bar{x}}\right)^{2}\bar{c}_{0}^{(0)} - \frac{\dot{\bar{T}}}{2\sqrt{2}\bar{T}}\bar{c}_{0}^{(1)}~.
\end{equation}

\section{Anomaly of heat rate}
\label{secS:heat}

\subsection{Heat-rate expression in terms of Brinkman's coefficient}

We now focus on the anomaly of heat rate that appears in the Langevin system with time-dependent temperatures. For an underdamped system, the average heat rate~\cite{Sekimoto1998} is given by
\begin{equation} \label{eqS:avg_heat_rate1}
    \langle \dot Q \rangle_\text{ud} = -\langle \gamma v^2 \rangle_\text{ud} + \langle v \circ \eta \rangle_\text{ud}~,
\end{equation} 
where $\circ$ denotes the Stratonovich product and $\langle \cdots \rangle_\text{ud}$ denotes the ensemble average with respect to the underdamped probability distribution $P_\text{ud}(x,v,t)$. Using the relation $\langle v \circ \eta \rangle_\text{ud} = \frac{\gamma T(t)}{m} = - \frac{\gamma T(t)}{m} \int dx \int dv~ v \partial_v P_{\rm ud}$ derived via stochastic calculus, Eq.~\eqref{eqS:avg_heat_rate1} can be reexpressed as
\begin{align}\label{dQ-Jirr-eqn}
    \langle \dot Q \rangle_\text{ud} &= \int_{-\infty}^{\infty} dx \int_{-\infty}^{\infty} dv~m v J_\text{ud}^\text{irr}(x,v,t) \nonumber \\
    &= \frac{m}{2} \int_{-\infty}^{\infty} dx \int_{-\infty}^{\infty} dv~ v^2 \left[-\partial_v J_\text{ud}^\text{irr}(x,v,t) \right], 
\end{align} 
where $J_\text{ud}^\text{irr}$ is the irreversible probability current associated with the underdamped system, given by 
\begin{equation}
    J_\text{ud}^\text{irr}(x,v,t) \equiv \left(-\frac{\gamma v}{m} - \frac{\gamma T(t)}{m^2} \partial_v \right)P_\text{ud}(x,v,t)~,
\end{equation}
and the integration by part is used for the second equality in Eq.~\eqref{dQ-Jirr-eqn}. The integrand in Eq.~\eqref{dQ-Jirr-eqn}, $-\partial_v J_\text{ud}^\text{irr}(x,v,t)$, can be further manipulated as
\begin{align} \label{eqS:integrand}
    -\partial_v J_\text{ud}^\text{irr}(x,v,t) = \mathcal{L}^{\rm irr} P_{\rm ud} = N e^{-\frac{\Phi}{2}} \bar{\mathcal{L}}^{\rm irr} \bar{P}_{\rm ud}.
\end{align}
By plugging Eqs.~\eqref{eqS:integrand} and \eqref{Pbar-exp} into Eq.~\eqref{dQ-Jirr-eqn} and using the eigenfunction relation $\bar{\mathcal{L}}^{\rm irr} \psi_n = -\varepsilon_n \psi_n$ with $\varepsilon_n = \gamma n/m$, the heat rate can be expressed as
\begin{align} \label{eqS:Q_rate_int}
    \langle \dot Q \rangle_\text{ud} 
    &= -\frac{\gamma}{2} \sum_{n=0}^\infty n c_n \int_{-\infty}^{\infty} dx \int_{-\infty}^{\infty} dv~ v^2 \psi_0 \psi_n~.
\end{align}
Using Eq.~\eqref{eqS:v_psi_n}, we can show that 
\begin{align} \label{eqS:v2_integration}
    \int_{-\infty}^{\infty} dv~ v^2 \psi_0 \psi_n = \frac{T(t)}{m} \left(  \delta_{n,0} + \sqrt{n} \delta_{n,2} \right).
\end{align}
Substituting Eq.~\eqref{eqS:v2_integration} into the integral in Eq.~\eqref{eqS:Q_rate_int}, we obtain
\begin{align}\label{dqdtud-eqn}
    \langle \dot Q \rangle_\text{ud} 
    &= -\frac{\sqrt{2}\gamma T(t)}{m} \int_{-\infty}^{\infty} dx~ c_2.
\end{align}
It is worth noting that, unlike the overdamped Fokker-Planck equation which includes only $c_1$ and $c_0$, the heat rate depends on the higher coefficient $c_2$. 

\subsection{High-viscosity regime: \texorpdfstring{$z=0$}{z=0}}

For the large-$\gamma$ regime, we use the expansion $c_2 = c_2^{(1)}/\gamma + c_2^{(2)}/\gamma^2 + O(\gamma^{-3})$, along with Eqs.~\eqref{c2n-eqns_a} and \eqref{c2n-eqns_b}, to obtain 
\begin{align} \label{eqS:heat_rate_terms}
    \langle \dot Q \rangle_\text{ud} =& \frac{\dot{T}}{2}\int_{-\infty}^{\infty} dx \left(c_0^{(0)} + \frac{1}{\gamma}c_0^{(1)}\right) -\frac{1}{\gamma} \int_{-\infty}^{\infty} dx ~f\left(f -T \partial_x \right)c_0^{(0)} \nonumber \\& +\frac{T}{\gamma} \int_{-\infty}^{\infty} dx ~ \partial_x \left( f - T\partial_x \right)c_0^{(0)} - \frac{m\ddot{T}}{4 \gamma}\int_{-\infty}^{\infty} dx~ c_0^{(0)} + O(\gamma^{-2})~.
\end{align}
In the large-$\gamma$ regime, $c_0^{(0)}+c_0^{(1)}/\gamma$ and $c_0^{(0)}$ in Eq.~\eqref{eqS:heat_rate_terms} can be substituted into $P_{\rm od}$. The second term on the right-hand side of Eq.~\eqref{eqS:heat_rate_terms} is identical to the average heat rate in the overdamped approximation, i.e.,
\begin{equation}
    \langle \dot Q \rangle_\text{od} = -\int_{-\infty}^{\infty} dx~ f J_\text{od}(x,t) = -\frac{1}{\gamma}\int_{-\infty}^{\infty} dx~f\left(f - T\partial_x \right)P_{\rm od}~,
\end{equation}
where $J_\text{od}(x,t)$ is the probability current in the overdamped approximation. Moreover, the third term on the right-hand side of Eq.~\eqref{eqS:heat_rate_terms} vanishes due to the natural boundary condition on $J_\text{od}$. Therefore, we finally arrive at
\begin{equation}\label{dQdt-ud-od-eqn}
    \langle \dot Q \rangle_\text{ud} = \frac{\dot{T}}{2} + \langle \dot Q \rangle_\text{od} - \frac{m \ddot{T}}{4 \gamma}~.
\end{equation} 
It is important to note that $\langle \dot Q \rangle_\text{od}$ and $\frac{m \ddot{T}}{4 \gamma}$ are of the same order in $\gamma$. In the previous literature, only the first term $\dot{T}/2$ of the right-hand side in Eq.~\eqref{dQdt-ud-od-eqn} is considered for the heat anomaly in the overdamped approximation. However, this is accurate only for processes with very slowly varying temperature. For temperature-varying processes with moderate speeds, the last term is also required for an accurate estimation of the heat anomaly in the large-$\gamma$ regime.

\subsection{Small-mass regime: \texorpdfstring{$z=1/2$}{z=1/2}}

Substituting the expansion~\eqref{cn-m-exp-all} in the small-$m$ regime into Eq.~\eqref{dqdtud-eqn} leads to
\begin{align}
    \langle \dot Q \rangle_\text{ud} \approx  -\sqrt{2}\gamma T(t) \int_{-\infty}^{\infty} dx~ c_2^{(2)} .
\end{align}
Using Eq.~\eqref{c22-m-exp}, we obtain
\begin{align} \label{eqS:small_mass_Q_ud}
    \langle \dot Q \rangle_\text{ud} =  \frac{\dot{T}}{2} \int_{-\infty}^{\infty} dx~c_0^{(0)}- \int_{-\infty}^{\infty} dx ~f J_\text{od}  +T \int_{-\infty}^{\infty} dx ~ \partial_x J_\text{od} .
\end{align}
The last term of Eq.~\eqref{eqS:small_mass_Q_ud} vanishes due to the boundary condition. Furthermore, $\int_{-\infty}^{\infty} dx~c_0^{(0)} \approx 1$ and $\langle \dot Q \rangle_{\rm od} =- \int_{-\infty}^{\infty} dx ~f J_\text{od}$. Thus, Eq.~\eqref{eqS:small_mass_Q_ud} simplifies to
\begin{equation}\label{dQdt-ud-od-eqn-2}
    \langle \dot Q \rangle_\text{ud} = \frac{\dot{T}}{2} + \langle \dot Q \rangle_\text{od}~.
\end{equation} 
We note that the heat anomalies in the small-$m$ and the large-$\gamma$ regimes are not identical; in the small-$m$ regime, the heat anomaly consists solely of the term $\dot{T}/2$.

\subsection{Intermediate regime: \texorpdfstring{$0<z<1/2$}{0<z<1/2}}

Equation~\eqref{dqdtud-eqn} can be rewritten in the following dimensionless form:
\begin{align} \label{dqdtudb-eqn}
    \langle \dot {\bar Q} \rangle_\text{ud} = - \sqrt{2}\tau \bar T \int_{-\infty}^{\infty} d\bar x~\bar c_2 \approx  - \sqrt{2}\tau \bar T \int_{-\infty}^{\infty} d\bar{x}~\left( \frac{1}{\tau}\bar{c}_2^{(0)} + \frac{1}{\tau^{2 - 2z}}\bar{c}_2^{(1)} \right)~,
\end{align}
where $\dot{\bar Q} \equiv (\tau_{\rm od}/T_0)\dot Q$. Substituting Eqs.~\eqref{eqS:bar_c2_0} and \eqref{c21-feqn} into Eq.~\eqref{dqdtudb-eqn}, we obtain
\begin{align} \label{eqS:bar_Q_z}
    \langle \dot {\bar Q} \rangle_\text{ud} \approx -\sqrt{2}\tau \bar T  \left[ -\frac{\dot{\bar{T}}}{2\sqrt{2}\bar{T}}\frac{1}{\tau} + \frac{\bar{T}}{\sqrt{2}}\frac{1}{\tau^{2 - 2z}} \int_{-\infty}^{\infty} d\bar{x} \left\{ \frac{\bar{f}}{\bar{T}} \left( \frac{\bar{f}}{\bar{T}} - \partial_{\bar{x}} \right) \bar{c}_0^{(0)} - \partial_{\bar{x}} \left( \frac{\bar{f}}{\bar{T}} - \partial_{\bar{x}} \right) \bar{c}_0^{(0)} \right\}\right]~,
\end{align}
where the normalization condition, $\int_{-\infty}^\infty d\bar x [\bar c_0^{(0)} + \bar c_0^{(1)}/\tau^{1-2z} ] \approx 1$, is used to obtain the first term on the right-hand side of Eq.~\eqref{eqS:bar_Q_z}. From the boundary condition, the last term on the right-hand side of Eq.~\eqref{eqS:bar_Q_z} vanishes. Thus, Eq.~\eqref{eqS:bar_Q_z} simplifies to
\begin{align} \label{eqS:bar_Q_z1}
    \langle \dot {\bar Q} \rangle_\text{ud} \approx \frac{\dot{\bar T}}{2} -  \frac{1}{\tau^{1 - 2z}} \int_{-\infty}^{\infty} d\bar{x} \bar f (\bar f - \bar T \partial_{\bar x})\bar c_0^{(0)}  ~.
\end{align}
If we revert the dimensionless variables in Eq.~\eqref{eqS:bar_Q_z1} to their original forms, the equation becomes
\begin{equation}
    \langle \dot Q \rangle_\text{ud} = \frac{\dot{T}}{2} + \langle \dot Q \rangle_\text{od}~.
\end{equation}
Therefore, $\mathcal{A}_Q$ for $0<z<1/2$ is identical to that for $z=1/2$. The key difference from the $z=1/2$ case is that the magnitude of $\langle \dot Q \rangle_\text{od}$ is lower than $\dot T/2$, i.e., $O(\langle \dot Q \rangle_\text{od})= \tau^{-1+2z}$, while they are the same for $z=1/2$.

\section{Anomaly of Entropy Production}

\subsection{EP-rate expression in terms of Brinkman's expansion}
The expression of EP rate for underdamped Langevin systems is given by
\begin{equation} \label{eqS:EPrate1}
      \langle \dot S_\text{tot} \rangle_\text{ud} = \frac{m^2}{\gamma T} \int_{-\infty}^{\infty} dx \int_{-\infty}^{\infty}dv ~\frac{(J_\text{ud}^{\rm irr})^2}{P_\text{ud}}~.
\end{equation}
Using the relation $P_{\rm ud} = \psi_0 \bar{P}_{\rm ud}$, the irreversible current $J_\text{ud}^{\rm irr}$ can be manipulated as
\begin{align} \label{eqS:J_ud_irr1}
    J_\text{ud}^{\rm irr} 
    &= -\frac{\gamma}{m}\left( v + \frac{T}{m} \partial_v \right) \psi_0 \bar{P}_{\rm ud} \nonumber \\
    &= -\frac{\gamma}{m}\left( v \psi_0 \bar{P}_{\rm ud}+ \frac{T}{m} (\partial_v \psi_0) \bar{P}_{\rm ud} + \frac{T}{m} \psi_0 \partial_v \bar{P}_{\rm ud} \right).
\end{align}
Here, $\partial_v \psi_0 = -\frac{mv}{2T} \psi_0$, and using the relation $H_n^\prime (x) = 2n H_{n-1} (x) $, $\partial_v \bar{P}_{\rm ud}$ can be calculated as
\begin{align}
    \partial_v \bar{P}_{\rm ud} &= \sum_{n=0}^{\infty} \frac{c_n}{\sqrt{2^n n!}} \left[  (\partial_v \psi_0)H_n\left(\sqrt{\frac{m}{2T}v}\right) + \psi_0 \partial_v H_n \left(\sqrt{\frac{m}{2T}v}\right)  \right] \nonumber \\
    &= -\frac{mv}{2T} \sum_{n=0}^\infty c_n \psi_n + \sqrt{\frac{m}{2T}} \sum_{n=1}^\infty \sqrt{2n} c_n \psi_{n-1}.
\end{align}
Therefore, putting these expressions for $\partial_v \psi_0$ and $\partial_v \bar{P}_{\rm ud}$ into Eq.~\eqref{eqS:J_ud_irr1}, we have
\begin{align} \label{eqS:J_ud_irr2}
    J_\text{ud}^{\rm irr} = -\frac{\gamma}{m} \sqrt{\frac{T}{m}} \psi_0 \sum_{n=0}^\infty \sqrt{n+1} c_{n+1} \psi_n ~.
\end{align}
Plugging Eq.~\eqref{eqS:J_ud_irr2} and the relation $P_{\rm ud} = \psi_0 \sum_{n=0}^\infty c_n \psi_n$ into Eq.~\eqref{eqS:EPrate1} leads to the expression for the EP rate in terms of Brinkman's expansion.

\subsection{High-viscosity regime: \texorpdfstring{$z=0$}{z=0}}
Retaining only the leading-order terms in the expression for the EP-rate in terms of Brinkman's expansion in the large-$\gamma$ regime, we obtain
\begin{align} \label{eqS:EPrate_mid}
    \langle \dot S_\text{tot} \rangle_\text{ud} 
    &= \frac{1}{m\gamma} \int_{-\infty}^{\infty} dx \int_{-\infty}^{\infty}dv ~ \frac{ \left[ c_1^{(1)} \psi_0 +\sqrt{2} c_2^{(1)} \psi_1   \right]^2}{c_0^{(0)}} + O(\gamma^{-2}) \nonumber \\
    &\approx \frac{1}{m\gamma} \int_{-\infty}^{\infty} dx \int_{-\infty}^{\infty}dv ~ \left[ \frac{\left(c_1^{(1)}\right)^2}{c_0^{(0)}} \psi_0^2 + \frac{2m \left( c_2^{(1)} \right)^2}{T c_0^{(0)}} v^2 \psi_0^2 + \frac{2\sqrt{2m} c_1^{(1)} c_2^{(1)}}{\sqrt{T} c_0^{(0)}} v\psi_0^2 \right] \nonumber \\
    &= \frac{1}{m\gamma} \int_{-\infty}^{\infty} dx \left[ \frac{\left(c_1^{(1)}\right)^2}{c_0^{(0)}} + \frac{2 \left( c_2^{(1)} \right)^2}{ c_0^{(0)}}  \right]~.
\end{align}
The integration results, $\int_{-\infty}^\infty dv~\psi_0^2 = 1$, $\int_{-\infty}^\infty dv~v\psi_0^2 = 0$, and $\int_{-\infty}^\infty dv~ v^2\psi_0^2 = T/m$, are used for the third equality in Eq~\eqref{eqS:EPrate_mid}. 
The first term on the right-hand side of Eq.~\eqref{eqS:EPrate_mid} corresponds to the EP rate in the overdamped approximation, as shown below:
\begin{equation}\label{Pi-od-eqn}
    \frac{1}{m \gamma}\int_{-\infty}^{\infty} dx~\frac{\left(c_1^{(1)}\right)^2}{c_0^{(0)}} = \frac{\gamma}{ T}\int_{-\infty}^{\infty}\!dx\,\frac{\left[\gamma^{-1} (f - T \partial_x )c_0^{(0)}\right]^2}{c_0^{(0)}} = \frac{\gamma}{ T}\int_{-\infty}^{\infty}\!dx\,\frac{J_{\rm od}^2}{P_{\rm od}} + O(\gamma^{-2}) \approx \langle \dot S_\text{tot} \rangle_\text{od}   ~.
\end{equation}	
The second term on the right-hand side of Eq.~\eqref{eqS:EPrate_mid} simplifies to
\begin{align}
    \frac{2}{m \gamma}\int_{-\infty}^{\infty} dx~\frac{\left(c_2^{(1)}\right)^2}{c_0^{(0)}} 
    &\approx \frac{m}{4 \gamma} \left(\frac{\dot{T}}{T}\right)^2\int_{-\infty}^{\infty}dx\,c_0^{(0)} \approx \frac{m}{4 \gamma} \left(\frac{\dot{T}}{T}\right)^2 ~.
\end{align}
Therefore, we finally reach the relation for the EP anomaly:
\begin{equation}
    \langle \dot S_\text{tot} \rangle_\text{ud}  
    =  \langle \dot S_\text{tot} \rangle_\text{od}  + \frac{m}{4 \gamma} \left(\frac{\dot{T}}{T}\right)^2~.
\end{equation}

\subsection{Small-mass regime: \texorpdfstring{$z=1/2$}{z=1/2}}

In the small-mass regime, the leading order contribution in the expression for the EP-rate in terms of Brinkman's expansion is
\begin{align} 
    \langle \dot S_\text{tot} \rangle_\text{ud} 
    \approx \gamma \int_{-\infty}^{\infty} dx \int_{-\infty}^{\infty}dv ~ \frac{ \left( c_1^{(1)} \psi_0 \right)^2}{c_0^{(0)}}  = \gamma \int_{-\infty}^{\infty} dx \left[ \frac{\left(c_1^{(1)}\right)^2}{c_0^{(0)}}  \right] \approx  \langle \dot S_\text{tot} \rangle_\text{od},
\end{align}
where the normalization condition, $\int_{-\infty}^\infty dv~\psi_0^2 = 1$, is used for the second equality. Therefore, we conclude that, unlike the large-$\gamma$ regime, the EP anomaly is absent in the small mass regime.

\subsection{Intermediate regime: \texorpdfstring{$0<z<1/2$}{0<z<1/2}}

The leading-order terms of $J_\text{ud}^{\rm irr}$ and $P_{\rm ud}$ in this regime are given by
\begin{align} \label{eqS:JandP}
    &J_\text{ud}^{\rm irr} = -\frac{\gamma}{m} \sqrt{\frac{T}{m}} \frac{\psi_0}{l_{\rm od}} \sum_{n=0}^\infty \sqrt{n+1} \bar c_{n+1} \psi_n \approx -\frac{\gamma}{m} \sqrt{\frac{T}{m}} \frac{\psi_0^2}{l_{\rm od}} \bar c_{1} ~, \nonumber\\
    &P_{\rm ud} = \frac{\psi_0}{l_{\rm od}} \sum_{n=0}^\infty \bar c_n \psi_n = \frac{\psi_0^2}{l_{\rm od}} \bar c_0\;.
\end{align}
Substituting Eq.~\eqref{eqS:JandP} into Eq.~\eqref{eqS:EPrate1} yields
\begin{equation} \label{eqS:EPrate1-2z}
      \left\langle \frac{d  S_{\rm tot}}{d\bar{t}}  \right\rangle_\text{ud} \approx \tau \int_{-\infty}^{\infty}d\bar{x} \frac{\bar{c}_1^2}{\bar{c}_0} 
      = \frac{1}{\tau^{1-2z}} \int_{-\infty}^{\infty} d\bar x \frac{\left[ \left(\bar{f} - \bar T\partial_{\bar{x}}\right) \bar{c}_0^{(0)} \right]^2}{\bar T \bar{c}_0^{(0)}}~,
\end{equation}
where Eqs.~\eqref{cb1-exp} and \eqref{eqA:FPbar-eqn1-0} are used for the second equality. Reverting the dimensionless variables in Eq.~\eqref{eqS:EPrate1-2z} to their original ones, we obtain
\begin{equation}
    \langle \dot S_\text{tot} \rangle_\text{ud} \approx \frac{\gamma}{T} \int_{-\infty}^{\infty} dx \frac{J_{\rm od}^2}{P_{\rm od}^2} = \langle \dot S_\text{tot} \rangle_\text{od}~.
\end{equation}
Therefore, no EP anomaly exists for $0<z<1/2$, and $O(\langle \dot S_\text{tot} \rangle_\text{od}) = \tau^{-1+2z}$.

\section{The origin of heat anomaly: Kinetic Energy Change}
In this section, we show that the heat anomaly in the overdamped approximation originates from the change in kinetic energy. In underdamped Langevin systems, $\langle v^2 \rangle_{\rm ud}$ is evaluated as
\begin{align} \label{eqS:v_square_1}
    \langle v^2 \rangle_{\rm ud} 
    = \int_{-\infty}^{\infty}dx \int_{-\infty}^{\infty} dv\, v^2 P_\text{ud}
    = \int_{-\infty}^{\infty}dx \int_{-\infty}^{\infty} dv\, v^2 \psi_0 \bar{P}_\text{ud}
    = \int_{-\infty}^{\infty}dx \sum_{n=0}^\infty c_n \int_{-\infty}^{\infty} dv\, v^2 \psi_0 \psi_n
    ~.
\end{align} 
Using Eqs.~\eqref{eqS:v2_integration} and \eqref{dqdtud-eqn}, Eq.~\eqref{eqS:v_square_1} can be further calculated as
\begin{align}
    \langle v^2 \rangle_{\rm ud} 
    = \frac{T}{m} + \frac{\sqrt{2}T}{m}\int_{-\infty}^{\infty}dx\; c_2 = \frac{T}{m}-\frac{\langle \dot Q \rangle_{\rm ud}}{\gamma}.
\end{align}
Thus, the mean kinetic energy is given by
\begin{align}
    \langle E_{\rm K}\rangle_{\rm ud} = \frac{m}{2}\langle v^2 \rangle_{\rm ud} \approx \frac{T}{2}-\frac{m}{2\gamma} \left(\langle \dot Q\rangle_{\rm od}+\mathcal{A}_Q \right) ~.
\end{align} 
Up to the order of $\langle \dot Q \rangle_{\rm od}$, the mean kinetic energy is approximated as
\begin{equation} \label{eqS:E_K_result}
\langle E_{\rm K}\rangle_{\rm ud} = 
    \begin{cases} 
        \frac{T}{2}  - \frac{m \dot{T}}{4 \gamma} & \text{for } z=0 \\ 
        \frac{T}{2} & \text{for } 0 < z \leq 1/2~.
    \end{cases}
\end{equation}
Therefore, the rate of mean kinetic energy is
\begin{equation}\label{dEkdt-eqn}
    \langle \dot E_{\rm K}\rangle = \mathcal{A}_Q~.
\end{equation}
Equation~\eqref{dEkdt-eqn} clearly shows that the heat anomaly arises from neglecting the kinetic energy change.

\section{Rapidly changing temperature}

In this section, we consider the case where the temperature changes rapidly as $\dot{T}/T \sim O(\gamma/m)$ or $\dot{\bar T} \sim O(\tau)$. Specifically, we set $\dot{\bar T}/\bar T = 2\tau a(t)$, where $a(t)$ is a dimensionless quantity of order $O(\tau^{0})$. From Eq.~\eqref{cbn-eqn_dimless_2}, the evolution equation for $\bar c_2$ can then be written as 
\begin{equation} \label{cn-rap-eqn}
    \partial_{\bar{t}} \bar{c}_2 = -\tau^z \sqrt{3\bar{T}(t)}\partial_{\bar{x}} \bar{c}_{3} + \tau^z \sqrt{2\bar{T}(t)}\left[ \frac{\bar{f}(x,\lambda_t)}{\bar{T}(t)} -\partial_{\bar{x}} \right] \bar{c}_{1} - 2\tau  \bar{c}_2 - \tau a(t) \left(2 \bar{c}_2 + \sqrt{2} \bar{c}_{0}  \right)~.
\end{equation}
We note that $\bar c_0$ is an $O(\tau^0)$ quantity, as it represents the marginal probability distribution. Therefore, the last term, $-\sqrt{2}\tau a(t)\bar c_{0}$, in Eq.~\eqref{cn-rap-eqn} is of order $O(\tau)$ and diverges in the large-$\tau$ limit. Since the probability distribution $P_{\rm ud}$ does not diverge in this regime, every $\bar c_n$ must remain finite. Then, the only terms that can compensate for this divergence are those related to $\bar c_2$, namely $-2\tau[1+a(t)] \bar c_2$. This implies that the coefficient $\bar c_2$ is of the same order as $\bar c_0$. Similarly, we can deduce that for an arbitrary $n$, the order of $\bar c_{n-2}$ is equal to the order of $\bar c_n$. Therefore, the expansion of the probability distribution in Eq.~\eqref{Pbar-exp} cannot be properly truncated, meaning the overdamped approximation fails when the temperature changes rapidly.

\section{Heat engine}

\begin{figure}[t]	
    \includegraphics[width=0.5\linewidth]{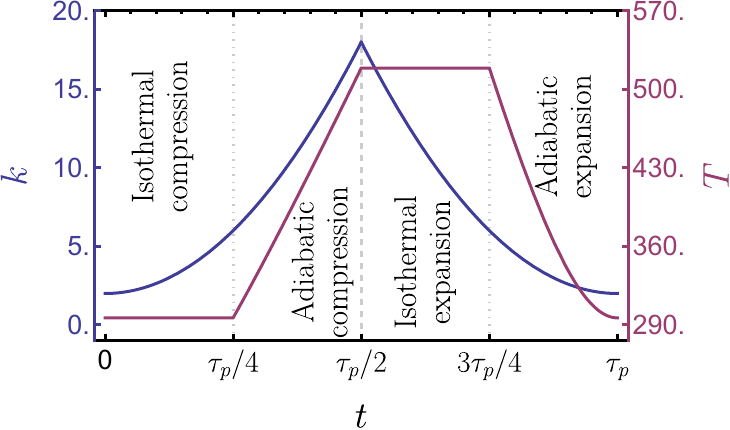}
    \caption{Plots of the time-dependent protocols of $k(t)$ and $T(t)$ as functions of time $t$. $\tau_{\rm p}$ denotes the period of the engine.}
    \label{Fig_S1}
\end{figure}

\begin{figure}[t]
    \includegraphics[width=\linewidth]{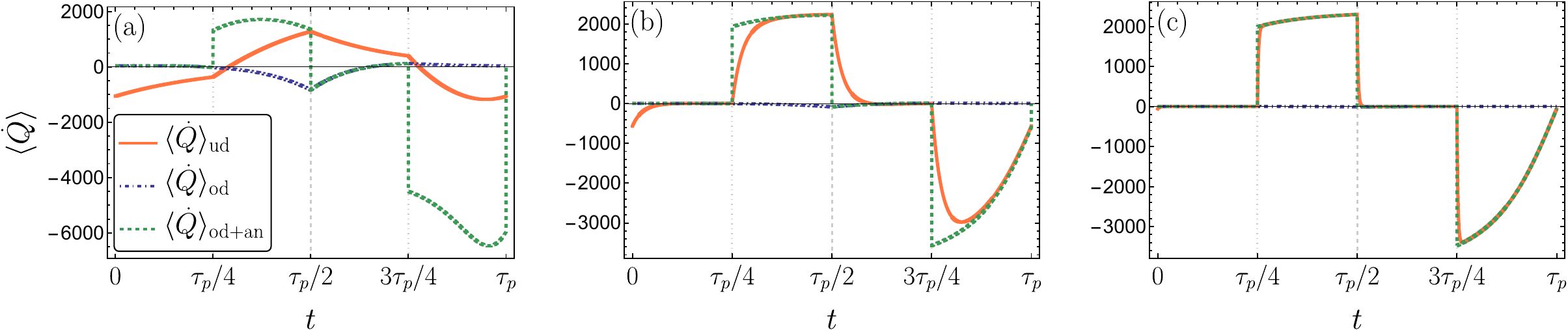}
    \caption{Plots of $\langle \dot Q \rangle_{\rm ud}$, $\langle \dot Q \rangle_{\rm od}$, and $\langle \dot Q \rangle_{\rm od+an}$ as functions of time $t$ for three different values of $\gamma$ in the Brownian Carnot engine model: (a) $\gamma = 10$, (b) $\gamma = 100$, and (c) $\gamma = 1000$.
    }  
     \label{Fig_S2}
\end{figure}

The time-dependent protocols for the temperature $T(t)$ and the stiffness of the harmonic potential $k(t)$ in the heat-engine model described in the main text are given by
\begin{align}
	&k(t) =  
	\begin{cases} 
		2+64t^2/\tau_{\rm p}^2 & 0\leq t < \tau_{\rm p}/2 \\
		2+64(\tau_{\rm p} - t)^2/\tau_{\rm p}^2 & \tau_{\rm p}/2\leq t < \tau_{\rm p}
	\end{cases}~,\\
	&T(t) = 
	\begin{cases} 
		T_c & 0\leq t < \tau_{\rm p}/4 \\
		T_c\sqrt{k(t)/k(\tau_{\rm p}/4)} & \tau_{\rm p}/4\leq t < \tau_{\rm p}/2\\
		T_h & \tau_{\rm p}/2\leq t < 3\tau_{\rm p}/4\\
		T_h\sqrt{k(t)/k(3\tau_{\rm p}/4)} & 3\tau_{\rm p}/4\leq t < \tau_{\rm p}
	\end{cases}~.
\end{align}
These protocols are visually represented in Fig.~\ref{Fig_S1}. 
To validate our theoretical results through numerical calculations, we simulate the engine model for three different values of $\gamma$, while keeping all other parameters identical to those in the main text.
Figure~\ref{Fig_S2} shows the plots of the three heat rates, $\langle \dot Q\rangle_{\rm ud}$, $\langle \dot Q\rangle_{\rm od}$, and $\langle \dot Q\rangle_{\rm od+an} = \langle \dot Q\rangle_{\rm od} + \mathcal{A}_Q$, as functions of time for $\gamma = 10,100$, and $1000$. In this calculation, $\mathcal{A}_Q $ for $z=0$ is used, as $\gamma$ increases to a large value in this setup. 
As shown in the figure, for $\gamma = 10$, the three heat rates do not coincide, as $\gamma$ is not large enough to ensure the validity of the overdamped approximation.
However, as $\gamma$ increases, $\langle \dot Q \rangle_{\rm od+an}$ approaches $\langle \dot Q \rangle_{\rm ud}$, whereas $\langle \dot Q \rangle_{\rm od}$ does not. $\langle \dot Q \rangle_{\rm od}$ coincides with $\langle \dot Q \rangle_{\rm ud}$ only during the isothermal process, where the temperature remains constant and no heat anomaly arises.

\section{Comparison of our method with the TAV method} 
\begin{figure}[t]
    \includegraphics[width=\textwidth]{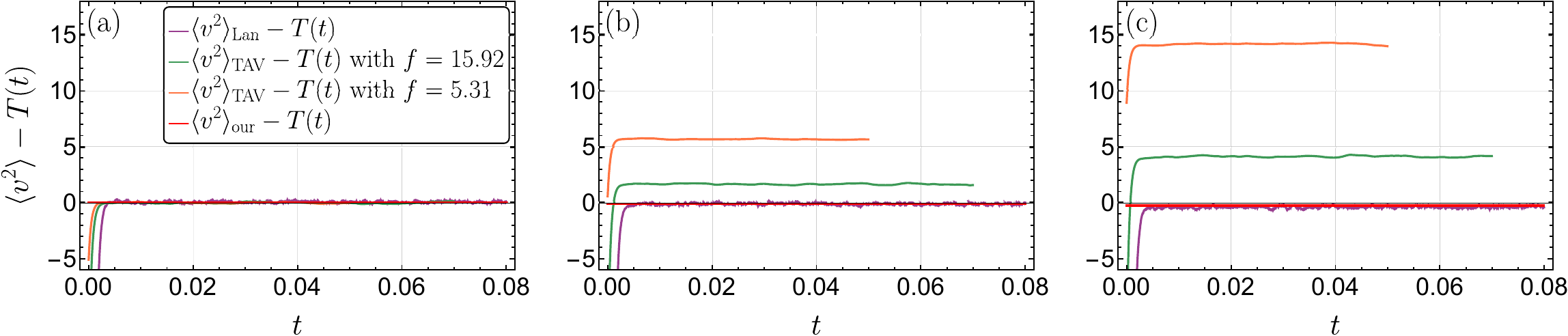}
    \caption{Comparison between $\langle v^2\rangle_{\rm TAV}$ and $\langle v^2\rangle_{\rm our}$. The figures show the time series of $\langle v^2\rangle_{\rm TAV} - T(t)$, $\langle v^2\rangle_{\rm our} - T(t)$, and $\langle v^2\rangle_{\rm Lan} - T(t)$. Two observation frequencies, $f = 5.31$ and $15.92$, are used for the TAV method. The temperature variation follows the form $T(t) = 300 + a t$, with $a = 0$, $400$, and $1000$ used for panels (a), (b), and (c), respectively.
    }
    \label{Fig_S3}
\end{figure}

We first summarize the time-averaged velocity (TAV) method~\cite{Roldan2014} used for comparison with our result in Fig.~\ref{Fig_S3}. Consider an underdamped Brownian particle trapped in a harmonic potential, whose dynamics are governed by the following Langevin equation:
\begin{equation}\label{lang_harm}
    \dot{x}_t = v_t, ~~m \dot{v}_t = - \gamma v_t-\kappa x_t + \eta_t ~,
\end{equation}
where $\eta_t$ represents the thermal noise. 

In a typical overdamped regime, the instantaneous velocity is not directly accessible. However, we can estimate the TAV, $\bar{v}_f(t)$, by measuring two consecutive positions with a sampling frequency $f = 1/(2\pi \Delta t)$ as follows:
\begin{equation}\label{vf-def}
    \overline{v}_{f}(t) \equiv \frac{1}{\Delta t}\int_{t}^{t+\Delta t}v(s)ds = \frac{x_{t+\Delta t} - x_t}{\Delta t},\quad 
\end{equation}
where $\Delta t$ is usually much larger than the relaxation time scale of the velocity. 
In Ref.~\cite{Roldan2014}, assuming the velocity is in equilibrium, it is shown that the mean-squared instantaneous velocity can be estimated from the mean square TAV using the following formula:
\begin{equation}\label{v2-vf-rel}
    \langle v_t^2\rangle=L(f)^{-1} \langle \overline{v}^2_f(t)\rangle~,
\end{equation}
where the function $L(f)$ is given by
\begin{equation}
    L(f)= 2f^2\left[ \frac{1}{f_0^2} + \frac{e^{-\frac{f_p}{2f}}}{f_1} \left(  \frac{e^{-f_1/f}}{f_p+2f_1} - \frac{e^{f_1/f}}{f_p-2f_1}  \right)\right]~
\end{equation}
with $f_p=\gamma/2\pi m, f_0=\sqrt{f_p f_\kappa}$, $f_\kappa=\kappa/2\pi\gamma$ and $f_1=\sqrt{f_p^2/4-f_0^2}$.

As an example, we consider the Langevin system described by Eq.~\eqref{lang_harm} with $\kappa = 1$, $m = 1$, and  $\gamma = 1000$. The temperature increases linearly over time as $T(t) = 300 + a t$, where $a$ quantifies the rate of temperature change. We evaluate the mean-square velocity $\langle v^2\rangle$ using three different approaches.
First, we obtain $\langle v^2\rangle_{\rm Lan}$ by solving the Langevin equation~\eqref{lang_harm} numerically, and then, averaging $3.2\times 10^7$ the resultant trajectories. 
Second, we apply the TAV method by computing $ \overline{v}^2_f(t)$ from Eq.~\eqref{vf-def} at sampling rates $f = 15.92$ and $5.31$, both at least an order of magnitude slower than the velocity relaxation rate $\gamma/m = 1000$, and estimating $\langle v^2 \rangle_{\rm TAV}$ from Eq.\eqref{v2-vf-rel}. Finally, we use our analytic expression~\eqref{eqS:E_K_result}, $\langle v^2 \rangle_{\rm our} = 2\langle E_{\rm K} \rangle/m$.

The results are presented in Fig.~\ref{Fig_S3}. For $a=0$, all three methods yield identical results. However, as $a$ increases, $\langle v^2 \rangle_{\rm TAV}$ deviates from $\langle v^2 \rangle_{\rm Lan}$, with larger deviation occurring for slower $f$. This results from the breakdown of the TAV method's assumption that velocity remains in equilibrium under rapidly varying temperature.
In contrast, $\langle v^2 \rangle_{\rm our}$ remains in excellent agreement with $\langle v^2 \rangle_{\rm Lan}$ even for relatively large $a$. This demonstrates the accuracy of our method in estimating the kinetic energy in overdamped systems.